\documentclass{ar3e}
\usepackage[dvips]{epsfig}
\normalsize
\newskip\zatskip \zatskip=0pt plus0pt minus0pt
\def\matth{\mathsurround=0pt}

\def\gsim{\mathrel{\mathpalette\atversim>}}
\def\atversim#1#2{\lower0.7ex\vbox{\baselineskip\zatskip\lineskip\zatskip
  \lineskiplimit
0pt\ialign{$\matth#1\hfil##\hfil$\crcr#2\crcr\sim\crcr}}}

\def\vereq#1#2{\lower3pt\vbox{\baselineskip1.5pt \lineskip1.5pt}}

\begin{document}
\input epsf.tex
\input psfig.sty
\jname{Submitted to Ann. Rev. Nucl. Part. Phys.}
\jyear{2004~~~~~~~~~~~~~~~~~~~~~~~~~~~EFI/04-03,~BNL-HET-04/02}
\vskip-0.2cm

\title{Physics Opportunities with a TeV Linear Collider}

\markboth{SD,MO}{Physics Opportunities with a TeV Linear Collider}

\author{Sally Dawson$^a$, Mark Oreglia$^b$
\affiliation{~$a$ Brookhaven National Laboratory, Upton, NY 11973\\
~$b$ The Enrico Fermi Institute, Chicago, IL 60637}}
\vskip-3em
\vskip-3em
\begin{abstract}

We discuss the physics motivations for building a 
500~GeV--1~TeV electron-positron linear collider. 
The state of the art collider technologies
and the physics-driven machine parameters are discussed. A survey of
some of
the phenomena well suited to study at a linear collider are described,
including Higgs bosons, supersymmetry, other extensions to the
Standard Model, and cosmology.

\end{abstract}
\vskip-0.20cm
\maketitle
\section{Introduction}
\label{intro}

\subsection{The Legacy of LEP and SLC}

High Energy Physics currently faces more opportunity than ever to address
significant questions about the structure of matter.  The 1967 Standard
Model (SM)~\cite{SM}
%
%
of electroweak unification is  consistent with experimental
observations to accuracies better than a part in $10^{3}$, and yet it now raises more questions
than it answers.  The origin of the symmetry breaking which
ultimately unifies electromagnetism with the weak force has still not
been found. 
There are theoretical blemishes in an otherwise elegant mathematical
model, such as the theoretical Higgs boson mass being unstable
to radiative corrections, requiring a suspicious fine-tuning
of Standard Model parameters in order to keep
the Higgs boson light~\cite{fine}. 
And then there is the fact that astrophysical observations suggest an abundance of a new
kind of matter. Many are convinced that we are poised for discoveries in
the near term which will significantly alter the physics
landscape.

\begin{figure}[htbp]
\centerline{
\epsfig{file=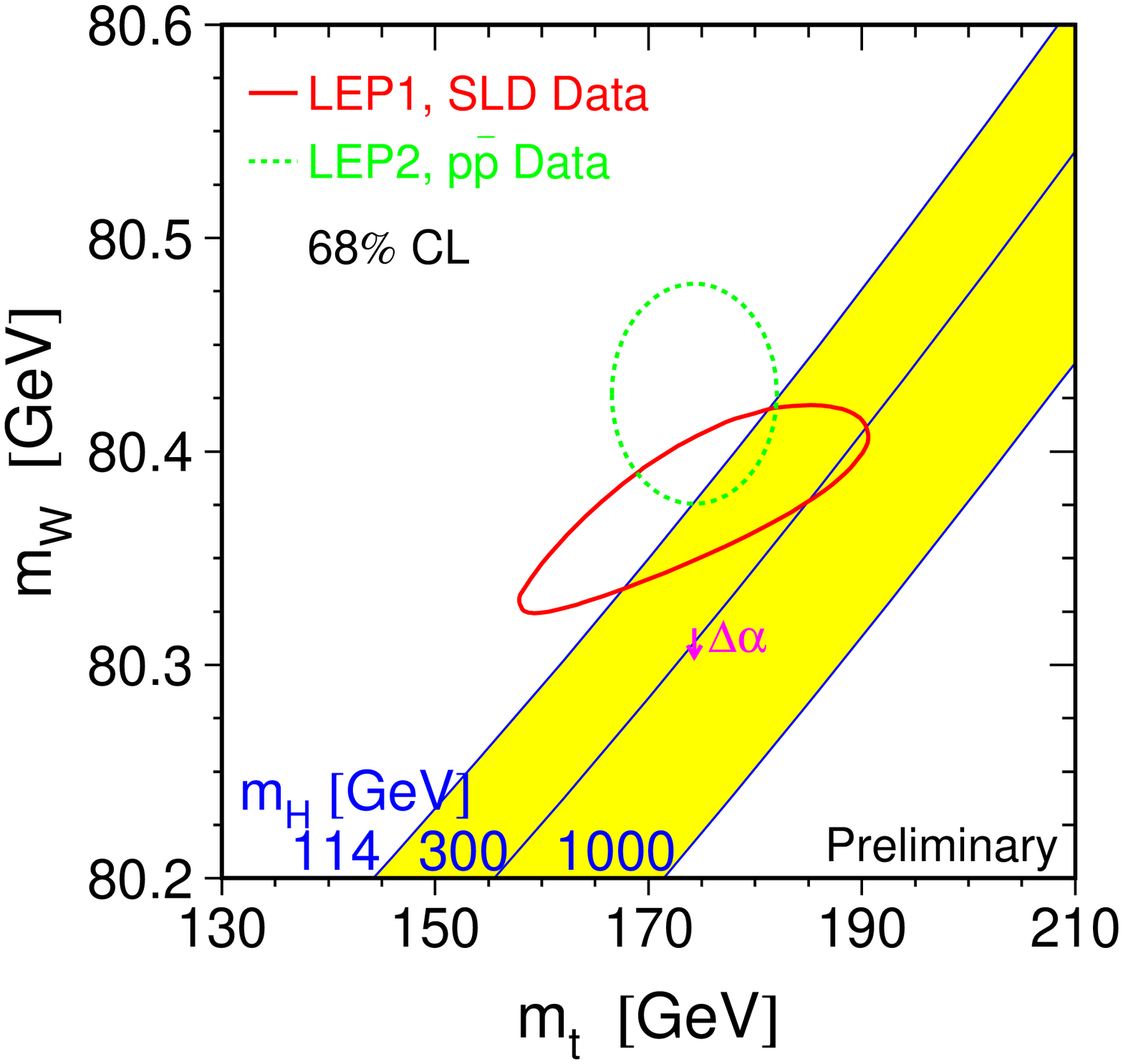,width=4.5in,angle=0}}
\vspace*{-.1cm}
\caption[]{\it 
Comparison of the indirect LEP-1 and SLD measurements of the W and top
masses (solid contour) and the direct LEP-2 and Tevatron measurements
(broken contour). Also shown is the Standard Model relationship for
the masses as a function of the Higgs boson mass.
The arrow labelled $\Delta\alpha$ shows the variation of this relation
if $\alpha(\mathrm{M}_{Z}^{2})$ is varied by one standard deviation.
From reference~\cite{LEPEWWG}.
}
\label{fig:mwmt}
\end{figure}

The precision measurements of electroweak processes by 
the Large Electron Positron collider (LEP) and 
the SLAC Linear Collider (SLC),
combined with the Tevatron's discovery of the top quark, 
have demonstrated to remarkable accuracy that a gauge theory 
involving a mechanism for spontaneous symmetry breaking (SSB)
describes physics nearly correctly at energies up to 200~GeV.
The energy scale for continued study of the electroweak sector is
fundamentally set by the value of the Fermi coupling constant, namely 300~GeV.
The LEP and SLD measurements depart from Standard Model
predictions for production cross sections and angular distributions at 
small but statistically significant levels.
These differences are generally consistent with modifications of the
gauge boson
propagators and decay vertices by a virtual particle or particles 
having properties similar to those of the Higgs boson in the
minimal version of the Standard Model~\cite{pt}. 
Fits to the combined electroweak data suggest\footnote{The limit
on the Higgs mass from fits to precision electroweak 
data  
is very dependent on the top quark mass; an increase in the top mass
by $\sim~30$~GeV increases the bound on the Higgs to roughly
$280~$GeV. These fits also assume the validity of the minimal Standard
Model with a single Higgs boson. The inclusion of new physics at the TeV
scale can significantly change this bound.}
that the mass of this particle is
in the $\sim$100-200~GeV region~\cite{LEPEW,LEPEWWG,LEPHiggs}.
%
%
The particle
could be the celebrated Higgs boson or some other object
responsible for symmetry breaking and particle masses.
The electroweak measurements are partially summarized in figure~\ref{fig:mwmt}
which shows the LEP Electroweak Working Group~\cite{LEPEWWG} 
%
%
compilation of W boson and top quark mass measurements. 
Superimposed on the figure are the
minimal Standard Model
predictions for the Higgs boson mass, starting from the
95\% confidence level lower limit of 114~GeV on its mass as obtained
from direct searches by the four LEP experiments~\cite{LEPHiggs}. 
%
%
The measurements indicate
that the Higgs boson
or some similar phenomenon is likely to occur at a mass
lower than approximately 200~GeV.  If the Higgs boson, or
some particle which plays the same role in the theory, does
not exist below around 1 TeV, then the self-interactions
of gauge bosons become strong and unitarity is violated~\cite{lqt}.
Hence on very general grounds, some new physics is expected
at an energy scale below 1 TeV.

Because of the rather well defined nature of the case
for new physics at the near-TeV scale, a strategy has been
formulated for High Energy Physics over the next 10--20 years. 
The Large Hadron Collider (LHC) will address many of the 
issues~\cite{lhc,atlastdr}, but has different
capabilities for discovering new physics 
than a linear collider because of the capability of a linear
collider for
high-precision measurements of masses and
missing energy and for polarized scattering.
Hence the community~\cite{HEPAP}
plans for a machine to address the
precision measurements -- a linear electron-positron collider (LC).  
The case for such a
machine has convinced the international community that serious
efforts are now needed to research and plan for the LC.
Studies for a linear electron-positron collider have existed for
nearly 20 years, an excellent review of LC physics appeared in
this journal 9 years ago~\cite{pm},
and there are a number of LC physics 
reviews~\cite{accomando,bagger,ellis0210052,concensus,snow2001,tesla,acfa}
%
%
%
%
%
%
%
 ... so what can we add to the case now?
The answer is: better understanding and new theoretical possibilities. 
The legacy of LEP and
SLC's precision mapping of the electroweak terrain, coupled with the
recent and exciting discoveries in astrophysics, have led to
extensions to the Standard Model and completely new theories for which
we need to again explore the suitability of the High Energy Physics 
tools of the coming years.   
Additionally, there have been significant advances in the LC
technology, which we attempt to summarize here.

\subsection{New Realities and Possibilities}

There are now several theoretical alternatives to the Standard
Model Higgs
boson, all of them aiming to cure the blemishes of the minimal model.
Perhaps the Higgs excitation is a fermion ``Cooper pair''
achieved by introducing a new set of strong interactions 
at the TeV scale~\cite{simmons,hills}.
In the ``Little Higgs'' theory, the Higgs boson is a composite
Nambu-Goldstone boson similar to the pion~\cite{littleh}.
New, heavy gauge bosons might exist which could help with the fine
tuning by means of loop corrections.
Perhaps there is no Higgs boson at all; divergences could be cancelled
by introducing an infinite tower of new particles in extra 
dimensions~\cite{nohiggs}.

Astrophysical measurements convincingly show that approximately 80\%
of the matter in the universe is non-baryonic and that
there is an astonishingly large amount of
dark energy~\cite{astrorev}. While the dark energy
issue continues to be baffling, the dark matter could well consist of
as yet undiscovered particles proposed to address some of the Standard
Model deficiencies. The nature of the dark matter suggests that it
interacts weakly; theoretical models such as supersymmetry (SUSY)
would furnish good candidates for the dark matter~\cite{susydm,eoss}.
The existence of supersymmetric particles would also offer a solution
to the Standard Model's fine-tuning problem.
Here there is also the exciting possibility that a precision
electroweak machine could be a window on Planck-scale physics
(e.g., if gravity mediates the supersymmetry breaking).

String theory now offers the possibility that there are more than
three spatial dimensions, and 
that there might be new particle excitations related to
the space-time structure. A low-background precision energy 
e$^+$e$^-$ linear collider
would see a distinct spectrum of narrow mass states arising from
excitations of the spatial structures~\cite{jo}.

The possibilities mentioned above are just a few of the new ideas
emerging to try to reconcile the Standard
Model with Planck-scale physics.  
While the theories are becoming numerous, they all have the
likelihood of new discoveries at the proposed LC energy.
The complex nature of many of the new theories will require
a low-background, precision energy collider to fully reveal the
theoretical structure.

\subsection{A Roadmap for High Energy Physics}

Currently, the Fermilab Tevatron is taking data on proton-antiproton
collisions at a center-of-mass energy of nearly 2~TeV.  While this
energy is in the range of some possible new phenomena, the backgrounds and
luminosity of the collider make it difficult to observe production of
particles with Higgs boson rates.  Experiments at this machine have
a better chance to discover larger cross section processes such as
supersymmetric particles. Tevatron data taking is expected to continue
for the remainder of this decade.

CERN's LHC is expected to be commissioned in 2007. This machine
collides protons at a center of mass energy of 14~TeV.  The energy and
luminosity are well matched for good identification of 
the Higgs boson and many candidate
particles occuring in
 extensions to the Standard Model; in particular, a SM
Higgs boson could be identified in 1 or 2-years of running for any mass
within the reasonable range of the theory~\cite{atlastdr}. 
 However, since the
center-of-mass energy of the fundamental collision is not known, and
the beams are not polarized, a number of high precision measurements
cannot be made. The LHC will likely produce the particles responsible for
symmetry breaking, but it will give us only partial knowledge of their
qualities.

The LC designs under consideration address precisely the limitations
of the LHC, though at a lower mass reach.  
But, more importantly, the LC capability of low-background precision
measurements opens a window to new physics.
In the best of worlds, LHC and LC data-taking would overlap so that
the two facilities could engage in feedback which would strengthen our
knowledge of nature. The case for combined running and complementarity
between the  LHC and a LC has been the topic of intense study~\cite{gw}.
An example of what is gained this way is realized by the complex chain
of the decay of supersymmetric particles.
 If supersymmetric particles exist, and they play the
expected role in curing the hierarchy problem, the LHC is likely to
produce many of them. 
Sleptons, in particular, are difficult to
identify cleanly at the LHC because 
they are produced with a small rate and large background with
missing energy in the decay. The LC
could provide these particle masses with much higher precision than
the LHC, 
and there are scenarios in which the
 LHC would not detect these sparticles
at all;
in such cases the mass information from the linear collider
would strengthen the LHC probe of 
the supersymmetric
parameter space at the highest energies~\cite{comp}.  In addition,
the capability of a linear collider to make precise measurements
of the couplings of new particles will be crucial
in order to verify that the 
 new particles correspond to an underlying supersymmetric model.
Some concurrent running of the two facilities suggests an LC
construction schedule.  LHC startup is expected in 2007, with real
physics output expected throughout 2008-2020.  Consequently, it would
be desirable to have the LC operational in 2015-17.  This would imply
initiation of LC construction by the end of this decade. 
A construction start at that time would be informed by the first LHC results.

\subsection{Physics-driven Accelerator Requirements}

The performance criteria enabling a LC to achieve the desired physics
measurements have recently been stated by the 
American Linear Collider Physics Group (ALCPG)~\cite{scopeA} 
and (with nearly
identical conclusions) by a committee established by the International
Linear Collider Steering Group~\cite{scopeI}.  
The major issues concern the energy,
luminosity, and beam qualities. 
%
%
%

It is reasonable to assume that the initial LC physics program would
have as its top goal copious production of the lowest mass Higgs
boson(s) or other particles in the 115-200~GeV mass range indicated by
LEP and SLD precision measurements~\cite{LEPEW,LEPEWWG,LEPHiggs}. 
The Standard Model production rates for the Higgs boson are lower
than those for most popular alternative physics scenarios, so it is
this Standard Model
 rate that sets the luminosity requirement for the machine.
Even if the current electroweak
precision measurements suggest a Higgs boson mass no higher
than approximately 200~GeV, the machine energy must be higher than
this mass for two reasons. 
Firstly, the 
important ``Higgstrahlung'' production mechanism (section~\ref{higgs}.1)
$\mathrm{e}^{+}\mathrm{e}^{-} \rightarrow \mathrm{Z} \mathrm{h}$
has a maximum yield for
a center-of-mass energy
$\sqrt{s}$ = 220 -- 340~GeV for Higgs masses in the 115 -- 200~GeV range. 
Second, detailed understanding of the particle properties will
require measuring the
 Higgs self-coupling, the hW$^+$W$^-$ 
coupling, and (for a higher mass Higgs) the
decay 
$\mathrm{h} \rightarrow \mathrm{t} \bar{\mathrm{t}}$~\cite{batt1,kd}.  
These processes suggest a baseline machine energy of 500~GeV, with the
capability to tune the energy in order to perform threshold scans and
some running at the Z peak.

The machine luminosity should be capable of producing enough Higgs
bosons to measure cross sections and couplings at the 5\% level
in order to distinguish among the 
models~\cite{snow2001,ghk,bd}
proposed to extend the Standard Model.  
The maximum Standard Model  Higgs production cross sections
(in association with a Z boson)
 vary from about 300 fb at $M_h=115$ GeV to about 70 fb
at  $M_h=200$ GeV.  Assuming that the luminosity scales linearly with 
$\sqrt{s}$, the
higher end of the mass range dictates the LC luminosity requirements,
namely an integrated luminosity of $500~ \mathrm{fb}^{-1}$.

Because precision electroweak measurements
and astrophysical arguments strongly indicate new physics in the
energy range of 100~GeV to beyond 1~TeV,
the linear collider will need to be upgraded to the TeV energy range
after the initial physics period. 
Exploration of the TeV energy region will be
crucial for our understanding of the Standard Model and for addressing
the indirect evidence of new physics.
The best place to study rare or low-rate Higgs decays 
is at higher energy using $\mathrm{W}^{+}\mathrm{W}^{-}$ fusion,
$\mathrm{e}^{+}\mathrm{e}^{-} \rightarrow \mathrm{h} 
\nu \bar{\nu}$
(section~\ref{higgs}.1 and reference~\cite{zerwas}).
This process has a cross section which increases 
with energy.
Running at an energy
near 1~TeV will allow
for good determination of small branching fractions such as
$\mathrm{h} \rightarrow \mu^{+} \mu^{-}$\cite{bdr}
with a fraction of the integrated luminosity it would take using
the $\mathrm{e}^+\mathrm{e^-}\rightarrow $Zh
 process at $\sqrt{s}=$500~GeV.
For light Higgs bosons, an important channel to measure is radiation
of a Higgs boson by a top quark in 
$\mathrm{e}^{+}\mathrm{e}^{-} \rightarrow \mathrm{t} \bar{\mathrm{t}}
\mathrm{h}$
(section~\ref{higgs}.2 and reference~\cite{snow2001}).
Other new physics processes benefit from the highest energy reach,
even at reduced luminosity.
For example, 
limits obtained 
on fermion compositeness and extra gauge bosons 
from the reaction
$\mathrm{e}^{+}\mathrm{e}^{-} \rightarrow \mathrm{f}\bar{\mathrm{f}}$ 
scale as $(\mathrm{s}^{2} L)^{0.25}$; 
for graviton exchange, the limits scale as
$(\mathrm{s}^{3} L)^{0.125}$~\cite{snow2001}, where $L$ is the
total integrated luminosity. 
In strong symmetry breaking scenarios one will want
to study final states with
$\mathrm{W}^{+}\mathrm{W}^{-}$,
ZZ, and $\mathrm{t} \bar{\mathrm{t}}$, which is most effectively done
at the TeV energy scale.

The LC will need to produce polarized beams in order to exploit the
parity violation intrinsic to the electroweak model
and to identify the quantum numbers of new particles. 
An added bonus of
polarized beams is the enhancement of signal 
reactions compared to most backgrounds.
Currently it is feasible to polarize the electron beam to better than
80\% as established by SLC,
 but positron polarization remains at the research and development stage.
Even if the machine were to operate with only polarized electrons,
much better sensitivity to SM parameters in 
$\mathrm{W}^{+}\mathrm{W}^{-}$ and $\mathrm{t}\bar{\mathrm{t}}$
production would be observed (section~\ref{other}.2), 
and it would allow for selection
of specific chirality in supersymmetric particle production
(section~\ref{SUSY}).
Electron polarization of 80\% or more is very effective in identifying
selectrons in supersymmetric theories,
where each lepton has two scalar partners which are produced preferentially
from polarized electron beams of opposite polarization.

Positron polarization increases the effective polarization, 
defined as
\begin{equation}
P_{eff}\equiv \frac{|P_{-}|+|P_{+}|}{1+|P_{-}P_{+}|},
\end{equation}
where $P_{+},P_{-}$ are the electron and positron polarizations.
This results in higher production rates of a number of
physics processes, including  Zhh and W$^+$W$^-$hh production~\cite{erler}.
Annihilation through a photon or virtual Z takes place from
the polarization states 
$\mathrm{e}^{-}_{\mathrm{L}}\mathrm{e}^{+}_{\mathrm{R}}$
and
$\mathrm{e}^{-}_{\mathrm{R}}\mathrm{e}^{+}_{\mathrm{L}}$
only.  By polarizing the positron
beam opposite to the electron beam, these channels are enhanced and
better defined.  For 80\% electron polarization, the introduction of 60\%
positron polarization leads to an effective polarization of 95\% and an
increase in the annihilation rate by a factor 1.5,
assuming no loss of intensity as a result of the
positron polarization~\cite{snow2001}.
Additionally, positron polarization aids in background suppression.  With an 
$\mathrm{e}^{-}_{\mathrm{R}}$ beam, 
introduction of a 60\% polarized positron beam leads to
a further factor of three background suppression
of W$^+$W$^-$ pairs.
Positron polarization can also be used to control signal and background
reactions involving $t$-channel exchange.   If the selectron appears at the
LC, positron polarization can select whether the partner of
the 
$\mathrm{e}^{+}_{\mathrm{L}}$ or $\mathrm{e}^{+}_{\mathrm{R}}$
is produced.  
Single $\mathrm{W}^{+}$ production is a background for some
analyses, and this can alse be controlled by using a polarized 
$\mathrm{e}^{+}_{\mathrm{L}}$ beam.
%

\subsection{Technologies and Issues for a Linear Collider}

\subsubsection{Accelerating Structures}

Electron storage rings cease to be effective for beam energies much
above 100~GeV because the power lost to synchrotron radiation grows as
$\mathrm{E}_{\mathrm{beam}}^{4}$. 
The total cost of electron synchrotrons scales roughly as 
$\mathrm{E}_{\mathrm{beam}}^{2}$,
whereas the scaling is linear for linear accelerating 
structures~\cite{accel}.
The only  practical
alternative for a next-generation electron accelerator
is to collide bunches accelerated in linear structures.
For cost reasons these linacs need to employ higher
acceleration gradients than used previously,
and this is facilitated by going to higher RF frequencies.  
The luminosity scales as (beam power)/(beam area),
with cost optimization requiring nanometer-scale beam area. 
Three acceleration
methods have been under development for well over a decade, two of
which look very promising for implementation in the short 
term~\cite{heuer}.

\begin{figure}[htbp]
\centerline{
\epsfig{file=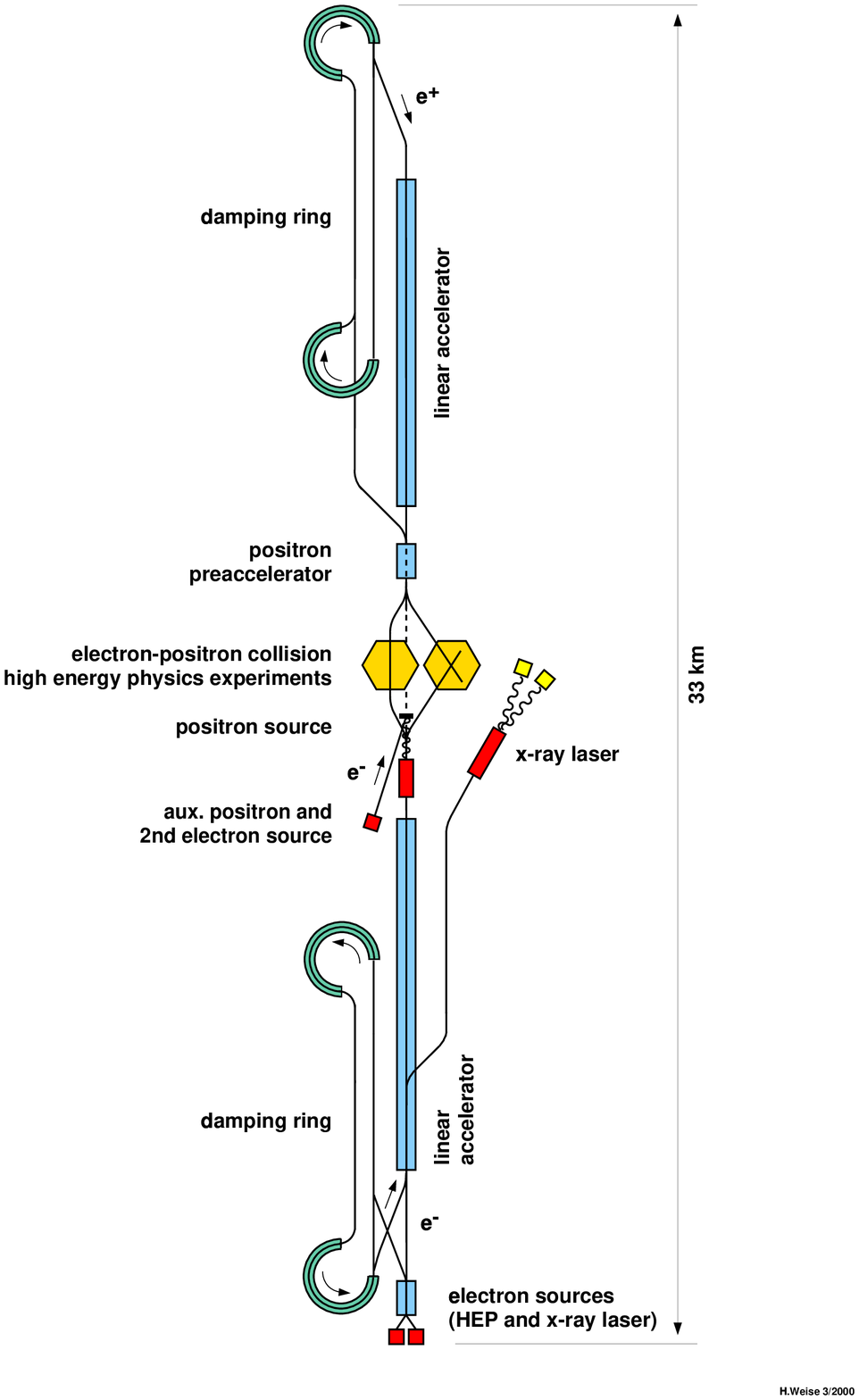,width=4.5in,angle=0}}
\vspace*{-.1cm}
\caption[]{\it 
Layout of the preliminary design for
the cold RF linear collider TESLA. 
The horizontal and vertical axes are not to the same scale,
so the collision angles appear much larger than they are in reality.
From reference~\cite{tesla}.}
\label{fig:TESLA}
\end{figure}

\begin{figure}[htbp]
\centerline{
\epsfig{file=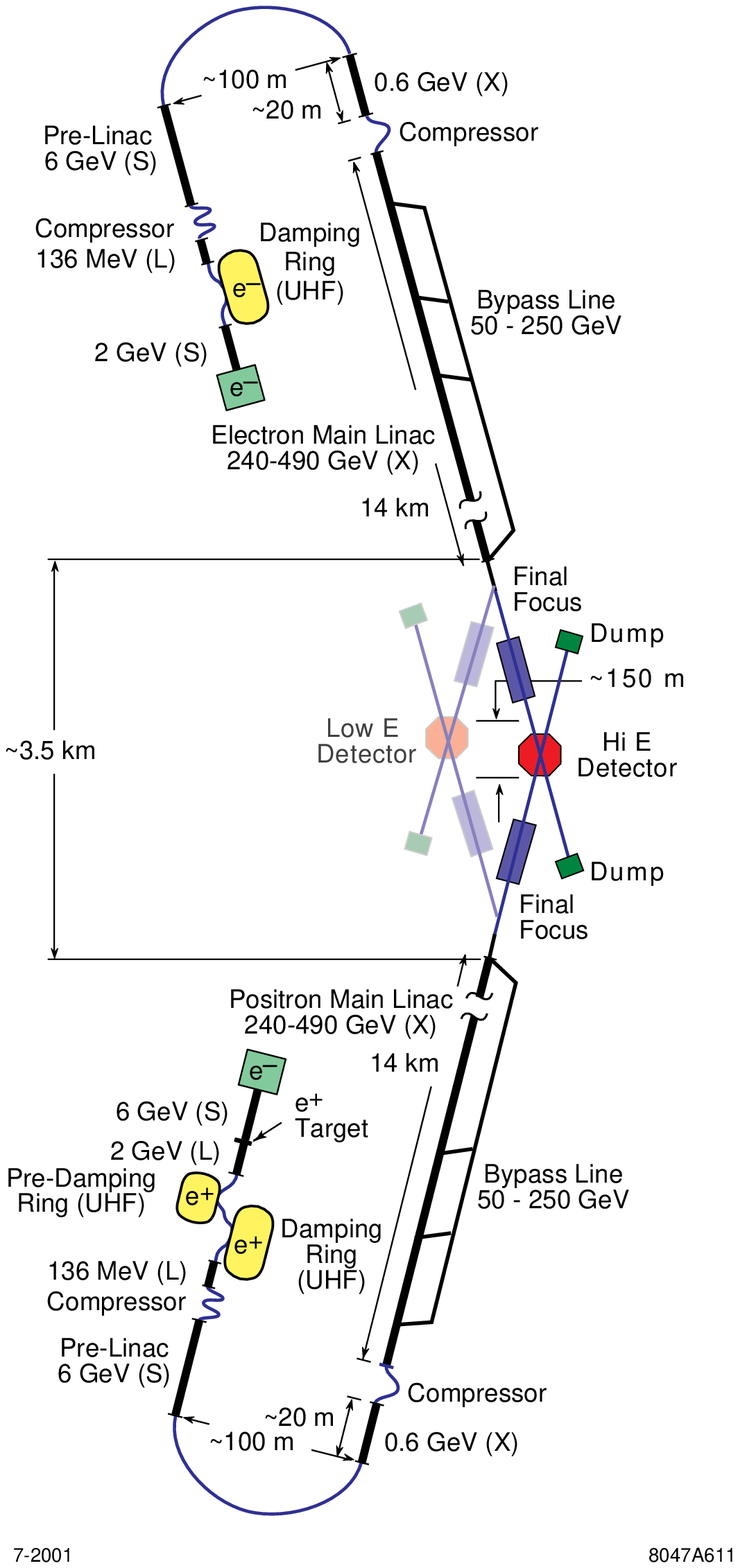,width=2.8in,angle=0}}
\vspace*{-.1cm}
\caption[]{\it 
Preliminary design for the
warm RF NLC  accelerator.
The horizontal and vertical axes are not to the same scale,
so the collision angles appear much larger than they are in reality.
From reference~\cite{warmRF}.}
\label{fig:NLC}
\end{figure}

\begin{itemize}

\item
Superconducting (or ``cold'') RF cavities are the main feature of the 
TESLA~\cite{tesla}
proposal. With superconducting structures, surface power density
must be limited and cryogenic losses increase as $\omega^{2}$ (where
$\omega$ is the RF frequency), 
so relatively long RF wavelength (1.4 GHz, or L-band)
must be employed. Accelerating structures capable of 35 MV/m have been
constructed and tested. 
This technology requires a relatively long
time between pulses because of the high-Q cavities.
The TESLA layout is shown in figure~\ref{fig:TESLA}~\cite{tesla}.

\item
Normal conducting (or ``warm'') RF cavities~\cite{warmRF,jlcref} 
%
%
have recently been
developed with beam-loaded gradients of 50 MV/m. 
In these structures stored energy scales as $\omega^{-2}$
and surface breakdown resistance as $\omega$,
so it is desirable to use the highest RF frequency possible.
The
higher X-band (11.4 GHz) frequency would be most cost effective;
a C-band (5.7 GHz) version is also under consideration. 
Like the superconducting case, the warm RF structures 
would be driven by conventional klystrons.
However, the warm RF scheme puts much greater stress on the klystrons,
requiring higher efficiency and much higher peak power.
The warm RF layout of the ``NLC'' scheme is shown in 
figure~\ref{fig:NLC}~\cite{warmRF}.

\item
The use of klystrons in L, C, and X band technologies is practical for beam energies up
to approximately 1~TeV, beyond which the power costs become
prohibitive. To reach the 3~TeV range, 
the Compact Linear Collider (CLIC)~\cite{CLIC} group
%
%
derives the RF power from deceleration of
a high-current, low-energy bunched beam.
It is thought that higher surface damage thresholds
can be achieved by using yet higher RF frequencies, so
CLIC is developing 30~GHz RF structures.
This technology could conceiveably achieve
gradients of 150~MV/m.  
Research and development of this challenging
technology is underway, but it is not expected to be 
demonstrated in the near term.
 
\end{itemize}

At the time this article was written, several issues with both the
warm and cold technology were identified by the Technical Review
Committee of the International Linear Collider Steering
Committee~\cite{ILCTRC}.
%
%
To grossly summarize the main issues raised in their exhaustive study,
it appears that both technologies are feasible. The two areas
most urgently identified for more study are 
RF damage in the warm cavities
and 
the feasibility of large-scale production of
cryogenic RF modules capable of 35~MV/m with acceptable
quenching and breakdown rates.
In addition, 
the extremely long damping rings needed by the cold
technology raise some concern.     
The choice of a technology will be made by the International LC
Steering Committee (ILCSC) after considering the recommendations
of its Technical Recommendation Panel~\cite{ILCTRP}.

\subsubsection{Beam Structure}

The superconducting and warm RF structures require markedly different
bunch structures.  The long time constants associated with
superconducting cavities lead to 950~$\mu$s pulses at a 5~Hz repetition
rate, while the warm RF has 267~ns pulses repeated at 120~Hz or
150~Hz.  The time between bunches in each pulse is 337~ns for
superconducting RF and 1.4~ns for warm RF.  
The detector design must accommodate these
differences in beam structure.
The long interbunch period for
cold RF allows detectors to read and clear between collisions, but the
overall pulse structure does not allow for easy electronics power
cycling to minimize cooling systems.  The warm RF time structure does
accommodate power cycling, but most detector electronics will not be
able to read individual bunch crossings, so there will be pileup and
duty factor issues.

\subsubsection{Beam Size and Collision Energy}

In order to achieve the luminosity required by the physics, the beams
must have nanometer scale transverse dimensions. This requirement
introduces special challenges for both the accelerator and the
detectors. The small bunch size at the collision point requires that
the accelerator maintain tight tolerances on the beam phase
space (or ``emittance'') through the several stages of transport, and that
beam jitter due to ground motion or other noise sources
be actively or passively corrected. 
Several groups have demonstrated the
feasibility of active feedback corrections for beam jitter,
though more work in this area is needed.

\begin{figure}[htbp]
\centerline{
\epsfig{file=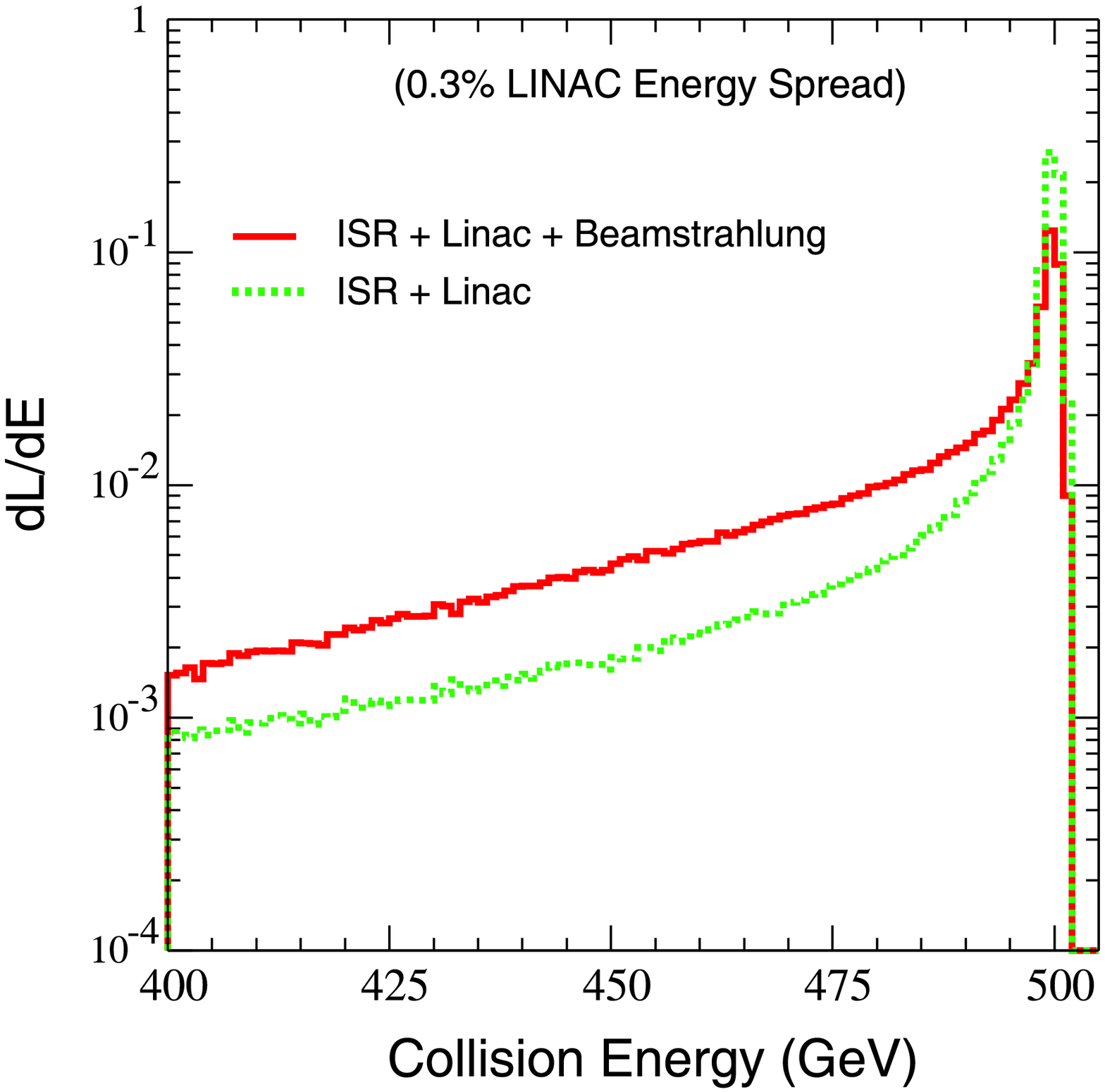,width=4.5in,angle=0}}
\vspace*{-.1cm}
\caption[]{\it 
Simulation of the luminosity spectrum expected for a 500~GeV warm RF machine,
including the effects of initial state radiation, beamstrahlung,
and linac energy spread of 0.3\%.
From reference~\cite{IPBI}.}
\label{fig:Ebeam}
\end{figure}

The extreme charge density at the collision point leads to new
phenomena -- and some problems -- for linear colliders, namely beam-beam
disruption, luminosity enhancement, photon emission and generation of
$\mathrm{e}^{+}\mathrm{e}^{-}$ pairs~\cite{accel}. 
For oppositely charged colliding beams, the bunches
focus each other with a highly nonlinear force.
This self-focussing leads to a enhancement in the luminosity by
a factor of 1.5 to 2~\cite{warmRF}.  However, the focussing also results in beams
which are more difficult to dispose of and backgrounds in the detector
from beam induced
synchrotron photons (known colloquially as
``beamstrahlung''~\cite{beamstrahlung}). 
The beamstrahlung results in a collision energy
spectrum with a very significant low energy tail,
as shown in figure~\ref{fig:Ebeam} (from reference~\cite{IPBI}.
Unlike the initial state radiation which can be accurately calculated,
the luminosity-weighted energy spectrum arises from energy variation
in the bunches caused by wake fields in the accelerating structures;
the spectrum therefore depends on the machine settings, 
and this will necessitate new types of beam instrumentation
to accurately measure the luminosity spectrum.
It is possible that the beams will have to cross at a small angle
in the interaction point in order to perform diagnostic measurements
of the beams, such as precise measurements of 
the beam energy, energy spread and polarization.  
Additionally, instrumentation for measuring the characteristics 
of the disrupted beam and beamstrahlung may improve the 
determination of the luminosity-weighted beam energy and polarization.  
The beam energy spread ($\sim$1000 ppm rms for TESLA) and luminosity-weighted 
beamstrahlung energy loss ($\sim$25000 ppm) are large compared 
to the 
desired precision for measuring the top mass, Higgs mass, W mass and 
the left-right decay product asymmetry ($\mathrm{A}_{\mathrm{LR}}$).  
The luminosity-weighted depolarization may be as 
large as 0.5\% at a 500 GeV LC, 
while the polarization
must be measured with
a precision of  0.1-0.25\%
for the physics program of measuring Standard Model asymmetries.
Extraction line polarization measurements may be needed to achieve
the desired precision~\cite{polarimeter}.  
Determining the luminosity-weighted energy and polarization 
from instrumentation that measures the average energy and 
polarization is non-trivial, and instrumentation is required
which still awaits development~\cite{IPBI}.  

%
%

\subsection{Collision Options}

The LC accelerator can be modified to facilitate
$\mathrm{e}^{-}\mathrm{e}^{-}$, 
$\gamma \gamma$, and $\mathrm{e}^{-}\gamma$ 
collisions.
The first process, M{\o}ller scattering, is the
reaction most sensitive to contact interactions and 
electron compositeness,
but there are other situations for which $\mathrm{e}^{-}\mathrm{e}^{-}$ scattering is important
to study~\cite{barger}.
For example, selectron pairs can be produced in $\mathrm{e}^{-}\mathrm{e}^{-}$ collisions by
$t$-channel exchange of neutralinos (section \ref{SUSY}), with Standard Model backgrounds extremely low.  
This process offers the best setting for high-precision measurements
of the selectron masses and the neutralino masses and mixing 
angles~\cite{fp,bfmpp,fmz}.

The $\mathrm{e}^{-}\mathrm{e}^{-}$ option also affords a low-background method
for generating $\gamma \gamma$ and $\mathrm{e}^{-}\gamma$ collisions by scattering
a high intensity laser pulse off the electron beams.
If the Higgs boson is in the 100-130~GeV mass range,
the photon scattering can be used to create a
`Higgs-factory' based on $s$-channel production~\cite{snow2001}. 
This facility would permit high precision measurement of the
$\mathrm{h} \rightarrow \gamma \gamma$
partial width, which is sensitive to new charged
particles such as the top squark, and also the top quark 
Yukawa coupling~\cite{gamgam}.  
The effective cross sections for producing Higgs and
Supersymmetric particles are as large or even larger than the
corresponding ones for $\mathrm{e}^{+}\mathrm{e}^{-}$, and
the capability of this machine to
produce linearly polarized beams not only can provide conclusive
information about the charge-parity (CP) nature of these particles, but
also will allow us to detect any CP admixture~\cite{snow2001}.
Certain rare decays 
such as 
$\mathrm{h} \rightarrow \gamma \gamma$, 
$\mathrm{h} \rightarrow \gamma \mathrm{Z}$, and 
$\mathrm{h} \rightarrow \mathrm{Z}\mathrm{Z}$ 
are highly sensitive to 
parameters of the Minimal Supersymmetric Standard Model (MSSM),
and could be cleanly measured with the photon collider.
A $\gamma \gamma$ collider could 
measure the spectrum and CP properties of the heavier Higgs bosons in
extended models
in $\tan{\beta}$ and $M_{A}$ regions that are not accessible to the 
$\mathrm{e}^{+}\mathrm{e}^{-}$ machine or the LHC.  
In a high
 energy $\mathrm{e}^{+}\mathrm{e}^{-}$ mode with unpolarized positrons,
the electron beam can be used to produce linearly polarized photons
to study the CP nature of the Higgs. 


While it will be necessary to occasionally operate the collider at the Z
resonance in order to calibrate the detector, one can evision a
program of high luminosity running at the ZZ and 
$\mathrm{W}^{+}\mathrm{W}^{-}$ thresholds 
in order to perform ultra-high precision measurements of electroweak
parameters\cite{giga,zfac}.  
Such a physics program would be worthwhile with approximately $10^{8}$
polarized Z's; 
at this level of statistics, the current uncertainty on the weak
mixing angle could be reduced by a factor of 5.
If positron polarization is possible, the measurements
at the Z resonance would be even more precise.

\subsection{Detector Challenges}

The detectors used to extract physics from a linear collider face
challenges from the accelerator architecture and
beam related backgrounds, 
as well as from the nature of the precision physics measurements needed. 
Because of the important physics with missing energy signatures, 
the detector will have to cover as much of the solid angle as
possible.
The detector will have to incorporate a channel for the 
disposal of the disrupted beams after collision, 
and elaborate masking must be devised to minimize
the beamstrahlung and other beam backgrounds in the detector,
especially near the beampipes. 
A large background from low energy
$\mathrm{e}^{+}\mathrm{e}^{-}$ pairs 
will require a high magnetic field for the central detector. 
High-rate two-photon physics processes 
will create large occupancy at low polar angles (with respect to the
beam axis), and therefore a high efficiency low-angle tagger will be needed.
The detector issues and emerging designs are summarized in references
\cite{schummdet,batdet,tesla_det,brau_sid}.
%
%
The physics phenomena driving the detector designs 
can be summarized as:
\begin{itemize}

\item excellent mass resolution (especially from the clean dimuon channel) to measure
recoil masses (Higgs bosons), kinematic edges (SUSY spectra), and
spectra indicative of other new physics; 

\item tagging of bottom and charm (Higgs sector, SUSY);

\item hadronic energy resolution capable of separating $W^{\pm}$ from
$Z$ jets, as well as Higgs boson and top quark decays (Higgs
properties, SUSY, other theories);

\item crack-free coverage of the solid angle down to low polar angles
for missing energy final states (SUSY, other theories). 

\end{itemize}

The flavor-tagging requires a
precision vertex tracker at very small radius. 
Highly pixellated semiconductor devices are the likely technology for the vertex
detector, and new semiconductor technologies promise to reduce the
long readout time required by older CCD devices.  
The physics requires impact parameter resolution of approximately
3-5 $\mu$m and a multiple scattering less than $~\sim 5 \mu\mathrm{m}/p$(GeV). 
The vertex detector will also need to be able to survive the large radiation
dose near the beampipe.

Particle tracking technologies under consideration 
include large-volume drift chambers or time projection chambers,
and smaller-radius solid-state detectors.
It is not yet clear that the necessary precision can be achieved with conventional
tracking-detector technologies.
The physics requirement on recoil mass resolution requires improvement
in the resolution from the current state of the art to 
$\delta p/p^{2} \sim 10^{-5}~\mathrm{GeV}^{-1}$.
Because of the large numbers of low energy pairs and multiple
collisions during the readout time, pattern recognition for tracking
is challenging. 
High efficiency tracking in the forward regions will be essential for missing
energy channels. 
There is also a potentially significant uncertainty on the dimuon
recoil mass measurement
from knowledge of the
luminosity-weighted energy distribution~\cite{snow2001}.
Since this distribution changes with machine running parameters,
instrumentation must be developed which can accurately measure this spectrum.  

\begin{figure}[htbp]
\centerline{
\epsfig{file=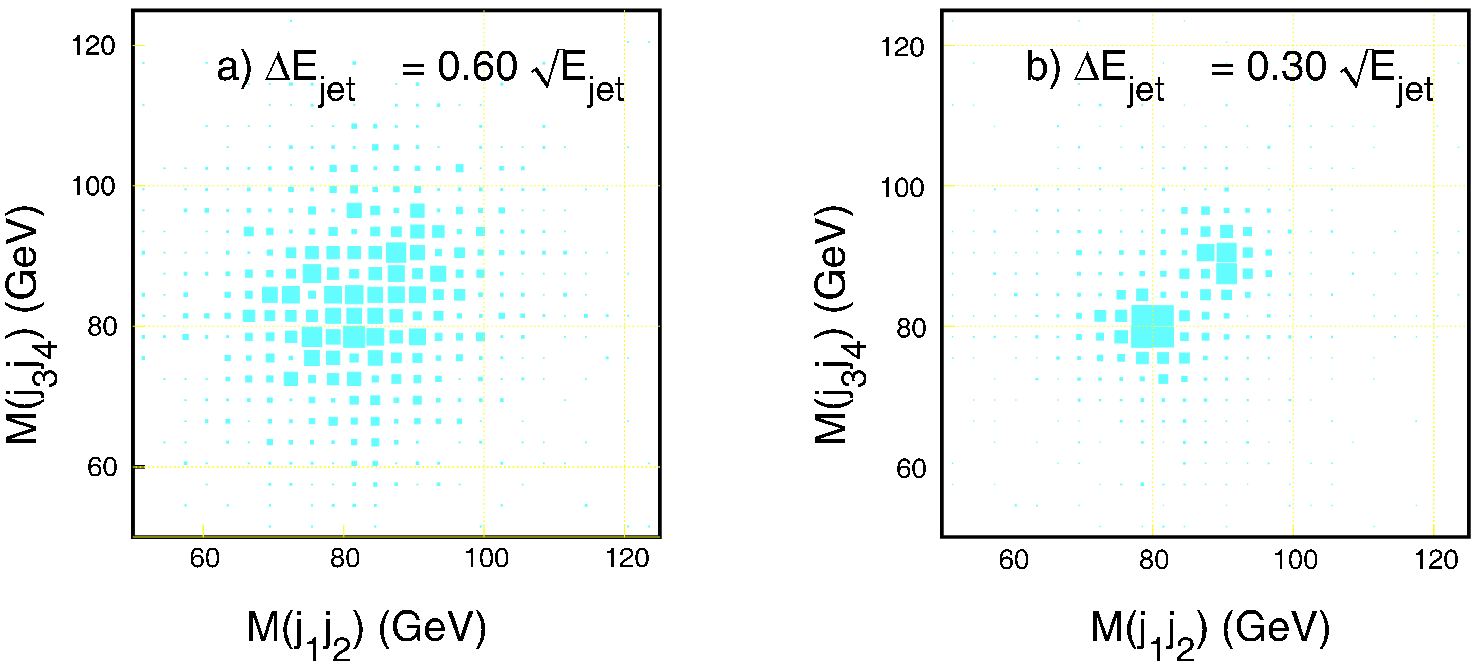,width=4.5in,angle=0}}
\vspace*{-.1cm}
\caption[]{\it 
Reconstructed dijet masses from simulation of 
$\mathrm{e}^{+}\mathrm{e}^{-} \rightarrow \nu \bar{\nu} ZZ$ and
$\mathrm{e}^{+}\mathrm{e}^{-} \rightarrow \nu \bar{\nu} WW$.
The jet energy resolution in a) is 60\%; in b) it is 30\%. 
From reference \cite{videau}.
}
\label{fig:videau}
\end{figure}

To discriminate Z decays
from $\mathrm{W}^{\pm}$ decays, the calorimeters
will need unprecedented jet energy resolution. 
Figure~\ref{fig:videau} shows simulations for reconstructed dijet
masses for 
$\mathrm{e}^{+}\mathrm{e}^{-} \rightarrow \nu \bar{\nu}$ ZZ or 
$\mathrm{e}^{+}\mathrm{e}^{-} \rightarrow \nu \bar{\nu} 
\mathrm{W}^{+}\mathrm{W}^{-}$.
The conclusion drawn is that the jet energy resolution must be 30\%/$\sqrt{E (GeV)}$ or
better~\cite{videau}.
Large diameter hadron calorimeters with the necessary granularity and resolution have
never before been built and energy-flow methods for final state
reconstruction are still under development. 
Energy flow techniques benefit from fine granularity in both the
electromagnetic and hadronic calorimeters,
and with sufficiently small granularity ($\sim$1 cm pads) it is conceivable that
the hadronic calorimeter information can be limited to ``digital''
mode -- that is, each pad is read out as ``on'' or ``off''.
It is not entirely clear that separate electromagnetic and hadronic
calorimeters, in contrast to a hybrid device, are the optimal choice. 

Differences in the bunch-timing structure of warm- and cold RF
machines have been mentioned earlier in section 1.5.2.
These different time structures will require that detectors
be optimized for the machine technology choice
from the standpoints of cooling, event pileup, and readout~\cite{brau_det}.
It will be possible to read and clear most detector technologies
between cold RF bunches, but this will likely be impossible for warm
RF (but the pileup does not appear to be a problem).
The exception to this is the vertex detector.
With cold RF, readout can currently be achieved in 50 $\mu$s,
which means there will be multiple readouts during the bunch train,
and RF noise can be a problem. 
The warm RF version would read out after the entire bunch train,
which avoids the RF noise pickup.
With either technology,
tracks will need to be time-stamped in order to reduce backgrounds
from overlapping collisions,
particularly from two-photon processes.
The longer inter-bunch spacing in a cold RF machine 
allows for easier time stamping.

\clearpage

\section{Higgs Bosons}
\label{higgs}
 
The experimental successes of the Large Electron Positron (LEP) collider
at CERN, the SLAC Linear Collider,  and the Tevatron Collider at Fermilab 
tell us that the standard $SU(2)\times U(1)$ gauge theory describes 
the interactions
of elementary particles
 at the $200~GeV$ energy scale to an astounding degree of precision.
The model, however, has one glaring flaw:  it cannot explain the source
of either gauge boson or fermion masses without the existence of an as
yet unseen  scalar particle.  This scalar particle
is dubbed  the Higgs boson.  In the minimal version
of the model, all particle masses are proportional to 
a single parameter, the Higgs vacuum expectation
value, $v$, which is determined by muon decay to be $v=246$~GeV.
The resulting gauge theory has a single unknown quantity, 
the mass of the Higgs boson, and all observables can be calculated
in terms of $M_h$.

The search for the Higgs boson is one of the most
 important components of the physics
program at both the CERN Large Hadron Collider and 
a future linear collider.  Finding the Higgs
boson and measuring its properties will give crucial 
insight into the underlying origin of mass and help to
complete our understanding of the $SU(2)\times U(1)$ 
gauge theory.

Finding a Higgs boson is only the
beginning, however,
 and inevitably  leads to more questions.  In order
to prove that the Higgs boson
 is the source of mass and to test the completeness
of the minimal Standard Model, three basic sets of measurements
are required:
\begin{itemize}
\item
The couplings of the Higgs boson to 
fermions and gauge bosons must be proportional to the
particle masses.  This
requires the measurement of Higgs boson branching ratios.
\item
The Higgs boson must be a spin-0 particle with positive charge
and parity.
\item
The Higgs boson must generate its own mass.  This requires measuring
the Higgs boson tri-linear and quartic self-couplings.
\end{itemize}

Should a fundamental Higgs boson not exist, alternative new physics
must provide the apparent effects described above (such as particle
masses, for example).  In section 4, we describe the role of
the linear collider in such a scenario.

\subsection{Producing the Higgs Boson at a Linear Collider}

A Higgs boson can be discovered at the LHC for any value of its 
mass up to the 1~TeV mass scale~\cite{atlastdr}. 
However, for most values of
the Higgs boson mass, the LHC is sensitive to only a 
subset of the possible Higgs decays and so is limited in its capability to 
obtain a complete picture of the Higgs boson couplings to 
matter~\cite{zep1}.
A linear collider can expand our knowledge of the Higgs boson properties
beyond the LHC capabilities~\cite{ghk}.

A linear collider  can copiously produce a Higgs
boson through
both the Higgstrahlung process,
\begin{equation}
\mathrm{e}^+\mathrm{e}^-\rightarrow \mathrm{Z} \mathrm{h}
\nonumber ,
\end{equation}   
and the vector boson fusion 
processes,
\begin{eqnarray}
\mathrm{e}^+\mathrm{e}^-\rightarrow & \mathrm{W}^+ 
\mathrm{W}^- \nu 
{\overline \nu}\rightarrow & \nu {\overline \nu} 
\mathrm{h}\nonumber \\
\mathrm{e}^+{e}^-\rightarrow & \mathrm{Z} \mathrm{Z}
\mathrm{e}^+\mathrm{e}^-
\rightarrow & \mathrm{e}^+\mathrm{e}^- \mathrm{h} . 
\end{eqnarray}
Both of these production mechanisms
are sensitive to the couplings of a Higgs boson to the W and Z
gauge bosons. 

\begin{figure}[H]
\centerline{
\epsfig{file=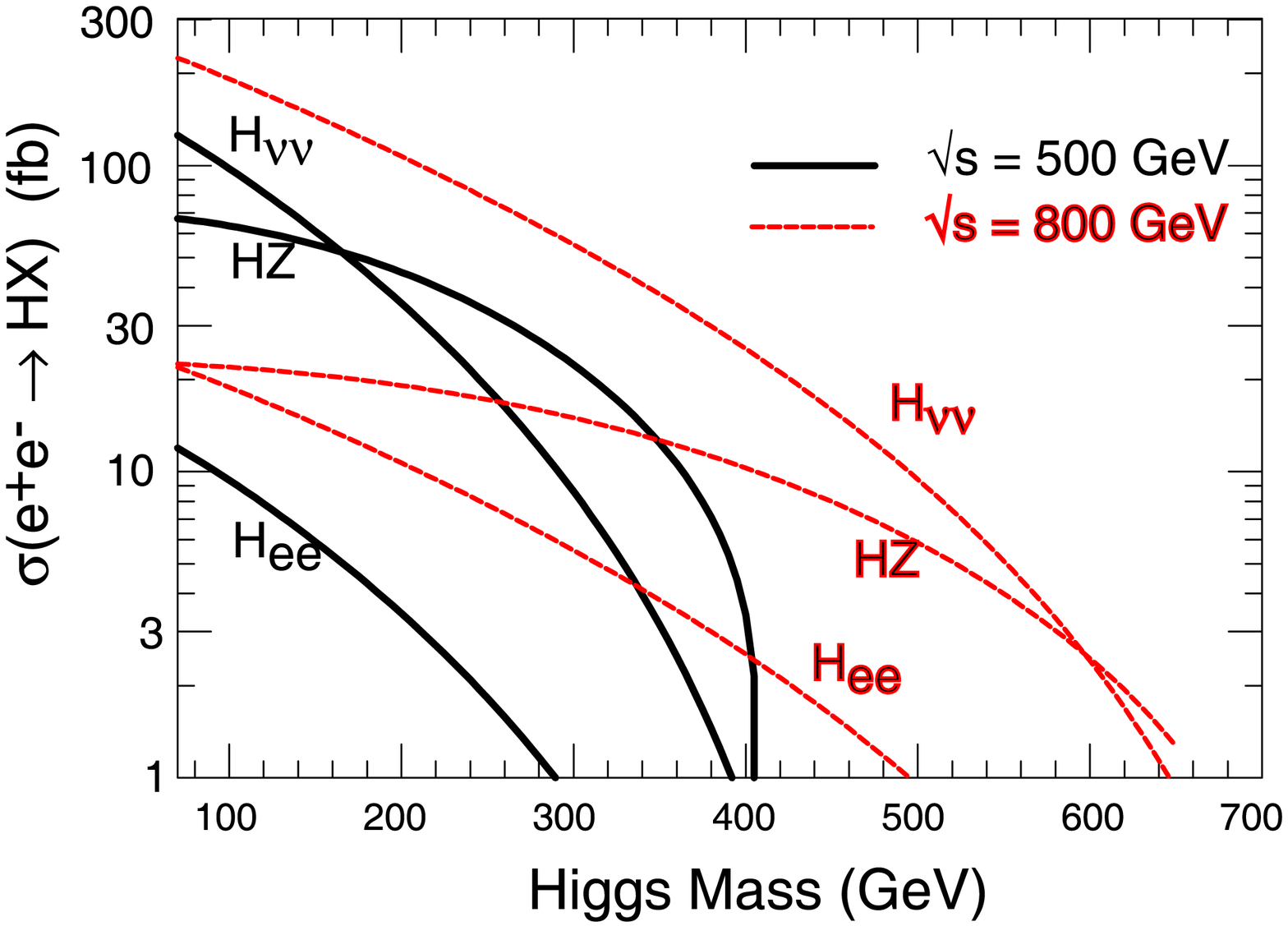,width=4.5in,angle=0}}
\vspace*{-.1cm}
\caption[]{\it 
  Cross section for the production of a Standard Model Higgs boson through
  Higgsstrahlung, $\mathrm{e}^+
\mathrm{e}^- \rightarrow \mathrm{Z}
\mathrm{h}$,  and in WW and ZZ
 fusion, $ \mathrm{e}^+\mathrm{e}^-
  \rightarrow \nu \overline{\nu}  \mathrm{h}$ and $ \mathrm{e}^+
\mathrm{e}^-
  \rightarrow  \mathrm{e}^+
\mathrm{e}^- \mathrm{h}$; solid curves: $\sqrt{s}
  = 500$ GeV, dashed curves: $\sqrt{s}= 800$ GeV. From reference~
 \cite{accomando}.}
\label{fig:hsigma}
\end{figure}

The Higgstrahlung process
reaches its maximum rate at an energy of
 $\sqrt{s}\sim M_Z+1.4 M_h-30$~GeV.
Measurements of the top quark mass, the W boson mass, and other electroweak
precision observables give an indirect limit on the Higgs boson
mass of $M_h<219$~GeV~\cite{LEPEWWG}, which suggests that the optimal
collider energy for observing the Higgs boson through Higgstrahlung
is $\sqrt{s}\sim 350-500$~GeV.  This is one of the major considerations
in determining the initial energy scale of a linear collider. On the
other hand,
the rate for Higgs production through vector boson fusion
 grows with energy and dominates the production rate
 at high energy. From figure~ \ref{fig:hsigma}, it is apparent that
the dominant production mechanism 
at high energy is $\mathrm{e}^+\mathrm{e}^-\rightarrow 
\mathrm{h}\nu{\overline{\nu}}$\footnote{For a heavy Higgs boson,
$M_h\gsim 500~GeV$, it is important to include the effects of the
Higgs boson width. In this case, calculating the on-shell cross section
and multiplying by the Higgs boson branching ratio 
 overestimates
the rate\cite{ctw}.}. 
The complete set of one-loop electroweak radiative corrections to both
Higgstrahlung and vector boson fusion are known,
 as are the strong ${\cal}O(\alpha_s)$ corrections~\cite{kniehl} and so
the predictions are on a firm theoretical basis.

\subsection{Measuring the Higgs Boson Couplings}

When the Higgs boson is produced in association with a Z boson,
 missing mass techniques
relate the Higgs boson 
mass to the initial state energy, $\sqrt{s}$,
and the energy of the detected Z boson, $E_{\mathrm{Z}}$, regardless
of how the Higgs boson decays,
\begin{equation}
M_h^2=s-2\sqrt{s} E_Z +M_Z^2~.
\end{equation}
Once the total cross section is measured by observing
the Z decay products, the various decays of the
Higgs boson can be reconstructed and the Higgs couplings
to the decay products  can be determined
unambiguously.   
There are two techniques for 
obtaining branching ratios from the $\mathrm{e}^+\mathrm{e}^-
\rightarrow 
\mathrm{Z} \mathrm{h}$ process~\cite{kd}.
  The first is to measure the total cross
section for $\mathrm{e}^+
\mathrm{e}^-\rightarrow \mathrm{Z} X$ and divide by the 
$\mathrm{e}^+\mathrm{e}^-\rightarrow
\mathrm{Z} \mathrm{h}$
 cross section obtained from the recoil mass technique.  This 
approach can also obtain the branching ratio for the decay of the Higgs
boson into an invisible mode, which occurs in some variants of the
Standard Model.  The second method is to obtain a sample of hZ events
and measure the fraction of events corresponding to 
$\mathrm{h}\rightarrow X$.

The Higgs boson branching ratios are definitive tests of the 
model~\cite{bd,brau_higgs,tesla_higgs}
and are expected to differ in extensions of the Standard 
Model~\cite{ghk}.  
For a light Higgs boson, $M_h < 160$~GeV,
the typical precision on the Higgs couplings to fermions  ranges
from $1-3\%$ for the coupling to the b quark, to $\sim
12\%$ for the couplings
to the charm quark and the $\tau$ lepton~\cite{tesla_higgs}.
The coupling to the muon is expected to
be poorly measured, $\sim 30\%$, due to its
small magnitude\cite{bdr}.  

The top quark is much heavier than the other quarks, 
$M_t\sim 175$~GeV, and
so plays a special role in many extensions of the Standard 
Model~\cite{simmons}.
The measurement of the Higgs boson coupling to the top is thus of particular
interest.  For $M_h<2 M_t$, this coupling is not accesssible from Higgs
decays, but instead can be measured through the associated production
process, $\mathrm{e}^+\mathrm{e}^-
\rightarrow \mathrm{t} \mathrm{{\overline t}} \mathrm{h}$.
  This mechanism suffers
from a small rate and requires high energy, $\sqrt{s}\sim 800-1000$~
GeV, and high luminosity.  An integrated luminosity of 
$L=1000~\mathrm{fb}^{-1}$ is
needed for a $10-20\%$ measurement of the Higgs-top quark
 coupling~\cite{juste}.



In the Standard Model, the Higgs boson self-couplings are dictated
 by the scalar
potential,
\begin{equation}
V={M_h^2\over 2} \mathrm{h}^2
+\lambda_3 v \mathrm{h}^3+{\lambda_4\over 4} \mathrm{h}^4,
\end{equation}
where  $\lambda_3=\lambda_4={M_h^2\over 2 v^2}$.
Verifying the relationship between $\lambda_3$ and $\lambda_4$
and the dependence of the self-couplings on the Higgs boson
mass  is a critical ingredient in demonstrating
that the Higgs boson is the source of its own mass.
Measuring $\lambda_3$ requires the
production of two Higgs bosons, and so the rate
at both a linear collider and the LHC
is very small and requires the highest possible luminosity.

At an energy of $\sqrt{s}=500$ ~GeV, 
an  $\mathrm{e}^+\mathrm{e}^-$ 
collider is sensitive to the production channel
$\mathrm{e}^+\mathrm{e}^-\rightarrow \mathrm{Z}
\mathrm{h}\mathrm{h}$.
  For $M_h<140$~GeV, the
dominant decay chain is $\mathrm{h}
\rightarrow \mathrm{b} {\overline \mathrm{b}}$ and there is
a high efficiency for identifying the b's recoiling from the $Z$
boson. This leads to a final state with four b quarks (from
the two Higgs boson decays).
The tri-linear Higgs self-coupling can be measured to roughly a $20\%$
accuracy for $120~GeV <  M_h < 140$~GeV with
an integrated luminosity of $1000 ~\mathrm{fb}^{-1}$ in
this channel and can definitively exclude 
$\lambda_3=0$~\cite{higgpot,lam4}.

In order to verify the structure of the scalar potential, it is necessary
to also measure the Higgs quartic coupling, $\lambda_4$.  Unfortunately,
this measurement remains elusive
at both the LHC and a linear collider because of the smallness of the rate
for triple Higgs production\cite{lam4}.

\subsection{Measuring the Higgs Boson Quantum Numbers}
 
\begin{figure}[htb]
\centerline{
\epsfig{file=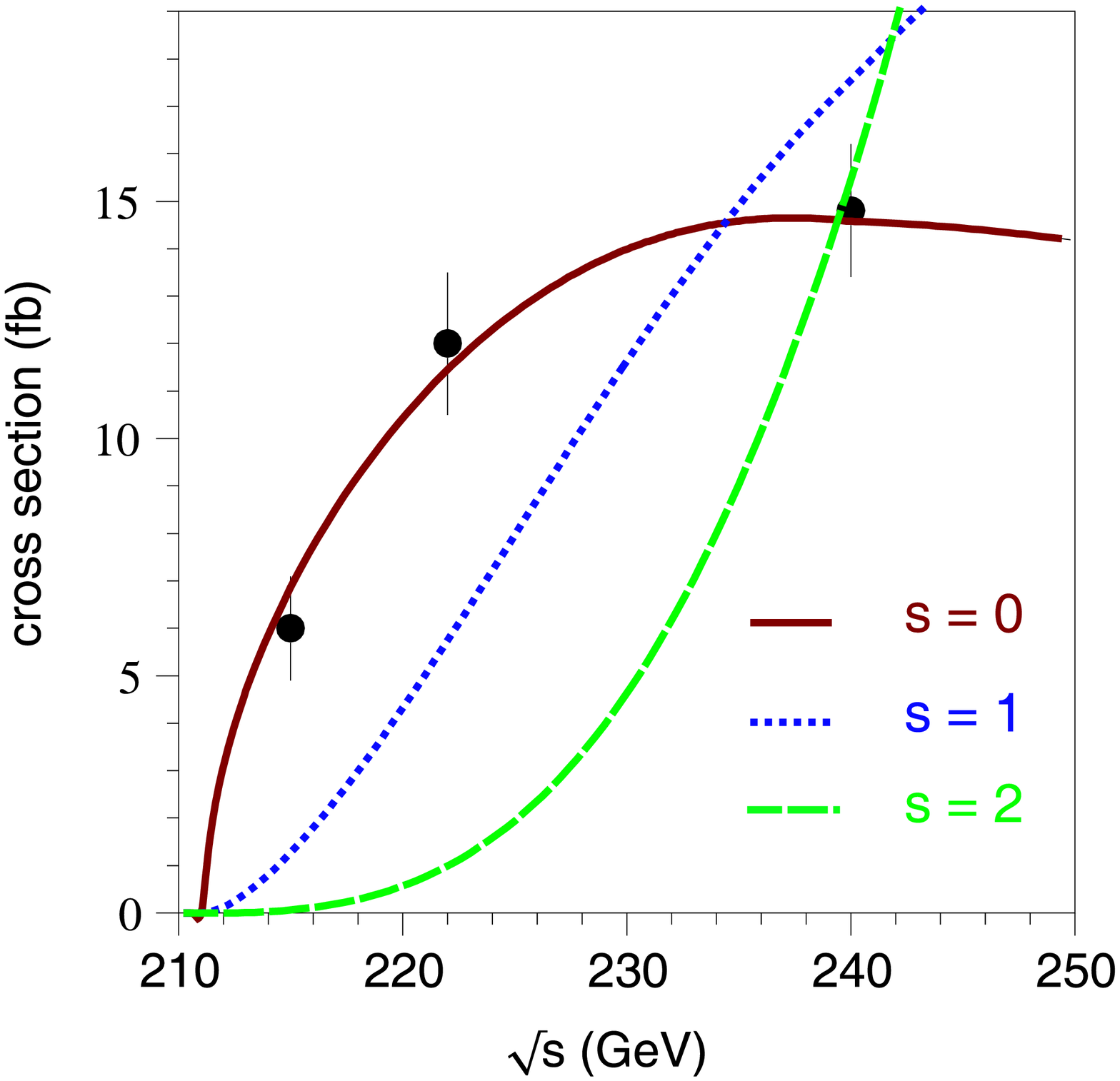,width=4.5in}}
\caption{\it The cross section for $\mathrm{e}^+
\mathrm{e}^-
\rightarrow \mathrm{Z}\mathrm{h}$ as a function of center-of-mass
energy for $m_h=120~GeV$ for a spin 0,1, or 2 Higgs
boson.  An integrated luminosity of $L=20~\mathrm{fb}^{-1}$
at each point is assumed.
From reference~\cite{dgl}.}
\label{fig:spin}
\end{figure}

One of the attractive features of a linear collider
is that it can
measure  the Higgs boson quantum numbers with few assumptions
about the underlying model. 
In order for the Higgs boson to have a vacuum expectation value, and
so be the source of mass, it must be a CP (charge/parity)
 even spin 0 particle. 
The angular distributions of both the Z and the h in the 
Higgstrahlung production
$\mathrm{e}^+
\mathrm{e}^-
\rightarrow \mathrm{Z}\mathrm{h}$ 
 are sensitive to the spin of the Higgs boson 
and scale as $d\sigma/d\cos\theta\sim \sin^2\theta$  for a $J^P=
0^+$ particle~\cite{miller}. 
The spin 
can also be determined by measuring the dependence of the cross section 
on the center-of-mass energy (see figure \ref{fig:spin}~\cite{dgl})
 and from the invariant
mass of the virtual Z boson in the decay, 
$h\rightarrow \mathrm{Z} \mathrm{Z^*}$ if
$M_h< 2 M_Z$~\cite{cmmz}.  The angular distribution of the decay
products in the $\mathrm{h}\rightarrow \mathrm{Z}
\mathrm{ Z^*}$ decay can also distinguish between
a Standard Model Higgs boson and a CP odd, $J^{PC}=0^{-+}$,
 pseudoscalar state.
These measurements are performed at the hZ threshold energy and
demand only a small luminosity, $L\sim 20 ~\mathrm{fb}^{-1}$.
Determining the Higgs boson quantum numbers from the decay
products does not
depend on the production mechanism and so can also be used to
determine the Higgs spin and charge conjugation in $\gamma\gamma$
collisions.  Equivalently, the 
observation of the decay $h\rightarrow \gamma \gamma$
rules out spin $1$ for the Higgs boson.  

\subsection{Higgs Spectroscopy in a Supersymmetric Model}
The model most often used for comparison with
the Standard Model is the minimal supersymmetric 
standard model (MSSM). The MSSM offers a rich phenomenology.
In the Higgs boson sector, there are  five Higgs bosons:  two
neutral scalars, $h^0$ and $H^0$, a pseudo-scalar, $A^0$, and two charged
scalars, $H^\pm$.  Because of the 
underlying supersymmetry, this is a
very predictive model. At the Born
 level, the Higgs sector has  only two free
parameters, which are typically taken to be the pseudoscalar mass, $M_A$,
and $\tan\beta$, the ratio of the neutral Higgs boson
vacuum expectation
values.   The lightest Higgs boson has an upper bound somewhere
around $M_h<130~GeV$~\cite{mhmax}, 
making it easily observable at an $\mathrm{e}^+\mathrm{e}^-$ 
collider with $\sqrt{s}=500$~GeV through the 
Higgstrahlung process\footnote{In non-minimal supersymmetric
models, the bound on the lightest neutral Higgs boson mass can
be increased to $150-200$~GeV by assuming that all couplings 
remain perturbative to the grand unification scale~\cite{nmssm}.}. 

\begin{figure}[htb]
\centerline{
\epsfig{file=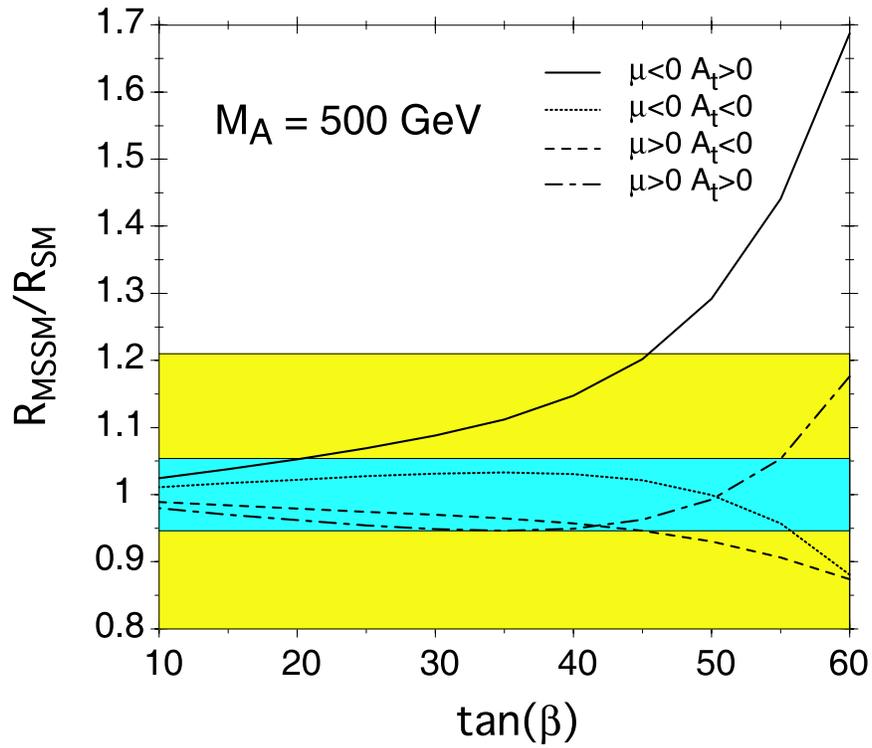,width=4.5in}}
\caption{\it Deviation of the ratio $R\equiv
\Gamma(h^0\rightarrow \mathrm{b}{\overline \mathrm{b}})/
\Gamma(h^0\rightarrow \tau{\overline \tau})$ from the Standard
Model prediction.  The expected experimental
uncertainty, $\sim~5\%$, is given by the inner
bands.
From reference~\cite{guasch}.}
\label{fig:gs}
\end{figure}

Precision measurements of the lightest neutral
Higgs boson branching ratios can give indirect
evidence for the existence of supersymmetry since these rates differ 
from those of the Standard Model .  Quantities such as $R\equiv
\Gamma(h^0\rightarrow \mathrm{b}{\overline \mathrm{b}})/
\Gamma(h^0\rightarrow \tau{\overline \tau})$ are particularly sensitive
to the parameters of the supersymmetric model~\cite{guasch}.  For large
values of $\tan \beta$, this ratio is sensitive to pseudoscalar masses
up to $600$~GeV, as can be seen in figure ~\ref{fig:gs}.
 A global fit to the 
expected precision for Higgs boson
decay rates at a $\sqrt{s}=350$~GeV collider
with an integrated luminosity of $L=1000~\mathrm{fb}^{-1}$ 
suggests that a linear
collider will be able to probe mass scales up
to around  $M_A\sim 600$~GeV for all values of $\tan\beta$~\cite{batt}.

As the pseudoscalar mass becomes large, the MSSM approaches a decoupling
limit and the Higgs sector looks much like that of the Standard
Model.  In this limit,
$A^0$, $H^0$ and the charged Higgs bosons, $H^\pm$ are almost degenerate
in mass and much heavier than the lightest neutral Higgs boson, while
the light $h^0$ couples to fermions and gauge bosons with a
similar magnitude as in  the Standard Model.  Therefore, in
 order to verify that
the theory is the MSSM, it is necessary to observe the heavier
Higgs bosons and measure their couplings. 

The dominant mechanisms for producing the heavier Higgs bosons
are pair production,
$\mathrm{e}^+\mathrm{e}^-
\rightarrow H^0 A^0$ and $e^+e^-\rightarrow H^+H^-$, and the 
corresponding mass
reach is approximately ${\sqrt{s}\over 2}$.  The production of
a single charged Higgs, $\mathrm{e}^+
\mathrm{e}^-\rightarrow \mathrm{b} {\overline \mathrm{t}} H^+$,
$\mathrm{e}^+\mathrm{e}^-\rightarrow \tau {\overline \nu}\tau H^+$, and 
$\mathrm{e}^+\mathrm{e}^-
\rightarrow W^\pm H^\mp$ can potentially extend the mass reach,
although the rates are rather small~\cite{single}.

\clearpage

\section{Supersymmetry}
\label{SUSY}

Supersymmetry is one of the most theoretically motivated extensions of the
Standard Model and tremendous attention has been devoted to possible
experimental tests~\cite{suprev,fn,snowsusy}. 
 It is a symmetry of space-time relating fermions
to bosons and is implemented by associating a new particle
differing by one-half unit of spin
with every known particle (these particles are collectively 
dubbed sparticles).   Supersymmetric  
models are very predictive and relate the couplings of
the new particles to those of the known particles.
This leads to rather generic predictions for the production
and decay rates of the sparticles in terms of the unknown sparticle masses.

The Standard Model suffers from a fine tuning problem when quantum
corrections to the Higgs boson mass are computed.  This problem is
cured in a supersymmetric model when the new particles associated with
the supersymmetry have masses on the TeV energy 
scale~\cite{drees}.  A linear collider
with a center-of-mass energy between $\sqrt{s}=500-1000$~GeV is thus
ideally suited to explore the spectrum of a supersymmetric model.

Observing the new particles associated with supersymmetry is a major
goal of both the LHC and a linear collider. 
 If supersymmetry is a symmetry at
the TeV energy scale, many of the sparticles 
will be discovered at the LHC~\cite{atlastdr}.
At a hadron collider, however, all of the kinematically allowed 
sparticles will be produced together, and it will be complicated
to untangle the pattern of sparticle masses and couplings~\cite{bhp}.
A lepton collider
has the capability to change its center-of-mass energy and so can
systematically explore the sparticle spectrum~\cite{snow}. 
 In addition, 
the scalar sparticles associated with the leptons will
be difficult for the LHC to observe and discovery may not occur
until the linear collider.  The exploration of the spectroscopy of
a supersymmetric model will occupy particle physicists 
at both the LHC and a linear collider for decades,
with complementary information gained from
the two machines.

A theory with unbroken supersymmetry has  particles and sparticles
of equal mass.   Once the supersymmetry is broken, the sparticles
obtain masses which differ from their corresponding particles.
Since supersymmetric particles are 
 not observed at the weak scale, supersymmetry must be
a broken symmetry.   Different mechanisms for breaking
the supersymmetry lead to varying patterns of sparticle masses
and so understanding the source of the
symmetry breaking will be a window to
unobserved higher energy scales and to understanding the source
of the supersymmetry breaking~\cite{suprev}.

Our goals for understanding supersymmetry include~\cite{fn}:
\begin{itemize}
\item
Find all the predicted sparticles.  
\item
Measure the sparticle couplings and quantum numbers.
\item
Measure the sparticle masses. 
\item
Understand the supersymmetry breaking scheme.
\end{itemize}

Supersymmetry associates a new scalar partner with each chiral fermion and so
each massive fermion has two scalar partners:  a scalar partner associated
with the left-handed fermion and a different scalar partner corresponding
to the right-handed fermion.  In a model with broken supersymmetry, these
scalar partners can have different masses.
Each gauge boson has a fermion partner associated with it:
the $SU(3)\times SU(2)\times U(1)$ gauge bosons have partners
called gauginos with masses $M_3, M_2$, and $M_1$ respectively.
The Higgs bosons also have fermion partners (Higgsinos). 
 The charged fermion
partners of the $W^\pm$ gauge bosons and the
fermion partners of the charged components of the  Higgs boson
doublets mix to form
particles called charginos, ${\tilde \chi}_i^\pm, i=1,2$. Similarly,
the fermion partners of the neutral Z and $\gamma$ gauge bosons,
and the neutral components of the
fermion partners of the Higgs bosons, mix to form
mass eigenstates called neutralinos, ${\tilde \chi}^0_i,i=1,4$.
A general model of supersymmetry breaking allows for 
arbitrary   masses for all of the new superpartners.

The most studied class of supersymmetric models contains a
symmetry called R-parity~\cite{dre}.  
This symmetry requires that supersymmetric
particles always be produced in pairs.  One consequence of R-parity
is that
the lightest supersymmetric particle is stable.  Since there are
stringent experimental limits on stable charged particles, the lightest
supersymmetric particle (LSP) is usually assumed to be the lightest
neutralino.  This particle also provides a viable candidate to explain
the dark matter in the universe~\cite{susydm}.

Supersymmetric models have been increasingly constrained 
by measurements of the branching ratio $B\rightarrow s\gamma$, the
muon anomalous magnetic moment, $a_\mu=(g-2)_\mu$, and the measurement
of the relic dark matter in the universe~\cite{susydm,eoss}.
In some classes of supersymmetric models,
these measurements can be interpreted as 
signals for supersymmetry at the few hundred GeV scale.

The linear collider can pair produce sparticles with masses
$M <\sqrt{s}/2$.  So the highest possible center-of-mass energy will
yield the maximum discovery potential for the new
particles predicted by supersymmetric models.
The tunable energy  of a linear collider 
allows for a search
strategy where the energy thresholds of the sparticles are
scanned sequentially. In addition, since the
sparticles
couple  differently to left- and
right-handed  fermions, polarization of the initial electrons
is effective in identifying specific production processes and reducing 
backgrounds~\cite{accomando,bagger}.

\subsection{sleptons}

In many models, the scalar partners of the electron (selectrons) and those
of the muon (smuons) are the lightest scalar particles.  
These scalars, collectively
termed sleptons, ${\tilde l}^\pm$, are produced through
$s-$ channel $\gamma$ and Z exchange,
\begin{equation}
\mathrm{e}^+
\mathrm{e}^-
\rightarrow \gamma, \mathrm{Z} \rightarrow {\tilde l}^+ {\tilde l}^- .
\end{equation}
Selectron pair production receives an additional $t-$ channel contribution
from  neutralino  exchange~\cite{fmpt}. 
Sleptons 
typically decay to either a neutralino, ${\tilde \chi}^0$, or 
to a chargino , ${\tilde \chi}^\pm$,
\begin{equation}
{\tilde l}^\pm
\rightarrow {\tilde \chi}^0 l,~ {\tilde \chi}^\pm \nu_l.
\end{equation}
If the slepton decays to the LSP (${\tilde \chi}^0_1$), 
the complete decay chain 
is 
\begin{equation}
\mathrm{e}^+\mathrm{e}^-\rightarrow {\tilde l}^+{\tilde l}^-
\rightarrow l^+l^- {\tilde \chi}^0_1{\tilde \chi}^0_1.
\end{equation}
The momentum of the leptons is precisely measured at a linear
collider, while  the 
neutralinos give missing transverse momentum in the detector.
 In models with $R$-parity 
conservation, the lightest neutralino is stable and
weakly interacting and  is not observed. 

The dominant Standard Model background  to slepton production and
decay to a neutralino is
\begin{equation}
\mathrm{e}^+\mathrm{e}^-\rightarrow 
\mathrm{W}^+\mathrm{W}^-\rightarrow
l^+l^{\prime~-}\nu_l {\overline \nu}_{l^\prime}.
\end{equation}
The $\mathrm{e}^+\mathrm{e}^-
\rightarrow \mathrm{W}^+\mathrm{W}^-$ 
Standard Model background is many times larger than the slepton signal,
but since it is strongly peaked in the forward direction, angular cuts
are effective in reducing this background.  
(Because the sleptons are scalar particles, they
 decay isotropically in their
rest frame and so they largely survive the cuts designed to eliminate  
the  W$^+$W$^-$ background.)

\begin{figure}[ht]
\centerline{
\epsfig{file=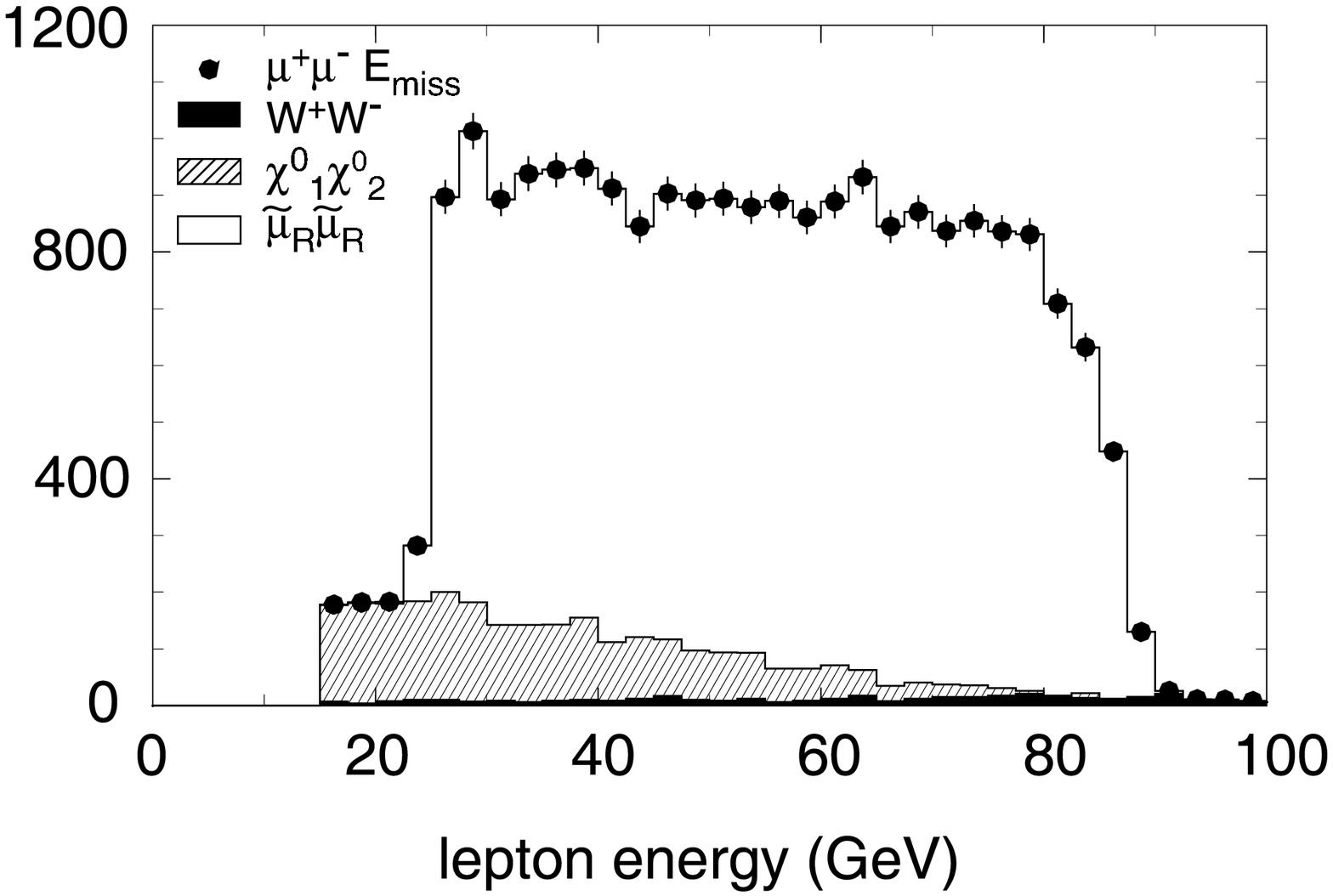,width=4.5in}}
\vspace*{-.1cm}
\caption[]{\it 
Signal from $\mathrm{e}^+
\mathrm{e}^-\rightarrow {\tilde \mu}_R^+{\tilde\mu}_R^-
\rightarrow \mu^+\mu^-{\tilde\chi}_1^0{\tilde\chi}_1^0$
and the dominant backgrounds for $m_{\tilde \mu_R}=132$~GeV
and $m_{\tilde \chi_1^0}=71.9$~GeV, $\sqrt{s}=320$~GeV, and
an integrated luminosity of $L=160~\mathrm{fb}^{-1}$. 
From reference~ \cite{mb}.} 
\label{fig:mur}
\end{figure}

The energy distribution of the outgoing leptons  from slepton 
decay is directly related
to the slepton and neutralino masses,
\begin{eqnarray}
m_{\tilde l}^2&=&{s E_{min}E_{max}\over (E_{min}+E_{max})^2}
\nonumber \\
1-{m_{{\tilde \chi}_0}^2\over m_{\tilde l}^2} &=&
{2\over \sqrt{s}}(E_{min}+E_{max}),
\nonumber \\
\end{eqnarray}
where $E_{max}$ and $E_{min}$ are the maximum and minimum allowed
lepton energies. 
The endpoints of the lepton energy spectrum can be clearly seen in
figure~  \ref{fig:mur}, where the distortion in the spectrum is the
result of initial state radiation~\cite{mb}.
 This technique yields measurements
of the slepton masses with accuracies in the few hundred $MeV$ range.

Slepton pair production provides an example of the effectiveness
of the polarization capabilities of a linear collider for reducing backgrounds.
The
$W$ bosons couple only to left-handed particles, so scattering
right-handed electrons
is extremely efficient for reducing the background, as demonstrated in
figure~ \ref{fig:mur}.
One of the definitive predictions of a supersymmetric theory is 
that of
the sparticle couplings.  A measurement of the absolute cross section
for slepton pair production measures the slepton couplings 
to $1-2\%$~\cite{pm}. 

The slepton masses can also be measured precisely from threshold
energy scans. This method is independent of the decay pattern
of the sleptons.
 Since the sleptons are scalars, the threshold
energy dependence scales like $\beta^3$, as compared with the $\beta$
dependence for  fermion pair production (where $\beta^2=1-4 M^2/s$ and $M$ is 
the mass of the produced particle). 
 Measurement of the slepton energy
dependence helps to verify that these particles are indeed scalars, as
required by supersymmetry.

A linear collider can also run in an $\mathrm{e}^-
\mathrm{e}^-$ mode.  This can potentially
yield more precise measurements of the 
slepton masses due to the steeper rise of the ${\tilde e^-}{\tilde e^-}$
cross sections at threshold as compared
to $\mathrm{e}^+\mathrm{e}^-$ 
collisions~\cite{fp}.   Furthermore, the background is
expected to be significantly smaller at an 
$\mathrm{e}^-\mathrm{e}^-$ collider.

\subsection{Charginos and Neutralinos}
A linear collider can pair produce both charginos and neutralinos,
\begin{eqnarray}
\mathrm{e}^+
\mathrm{e}^-&\rightarrow& {\tilde \chi}^\pm_i {\tilde \chi}_i^\mp,\quad
i=1,2\\
\mathrm{e}^+
\mathrm{e}^-&\rightarrow& {\tilde \chi}^0_i {\tilde \chi}^0_i,\quad i=1,4.
\end{eqnarray}
In most models, the lightest neutralino is stable and non-interacting, so
direct pair production of the
LSP  will not be observable.  The heavier neutralinos
will decay to states including  the lightest neutralino.  
Here we focus on chargino
pair production as an example of the capabilities of a linear
collider.

Chargino production occurs through $s-$ channel Z exchange and $t-$
channel sneutrino exchange.  The charginos decay to final
states with multiple jets and missing transverse energy from the
unobserved neutralino.  The dominant backgrounds are gauge boson pair
production, W$^+$W$^-$ and ZZ, which can be minimized using polarization.

Measurement of the chargino mass probes the entries of the chargino mass
matrix: $\mu$, $M_2$, $\tan\beta$, and the sneutrino mass which enters
into the production cross section. ($\mu$ is a Higgs mixing parameter, which
is required in order to have electroweak symmetry breaking, $M_2$ is
the mass of the fermionic partner of the W, and $\tan\beta$ is the
ratio of the vacuum expectation values of the neutral Higgs bosons.)
 The chargino mass matrix is
\begin{equation}
m_{\tilde \chi^\pm}=\left( \begin{array}{ll}
 M_2 & \sqrt{2} M_W \sin\beta
\\
\sqrt{2} M_W \cos \beta & \mu\end{array}\right).
\end{equation}
The off-diagonal terms are often much smaller than $M_2$ and $\mid
\mu\mid$ and in this scenario
 $m_{\tilde \chi^\pm}$ directly measures $M_2$
or $\mu$. 

The chargino mass can be 
measured through the kinematic endpoints of the decay products and also
through a threshold scan.  The ability of a linear collider to make
an energy scan and to have beam polarization allows a series of
measurements,
\begin{equation}
\sigma_L, \sigma_R,  A^{FB}_L, A^{FB}_R, m_{\tilde \chi^\pm},
\end{equation}
where the $L,R$ subscripts denote left- and right- polarized electron
beams, and $A^{FB}$ is the forward backward asymmetry.
$\sigma_R$ contains only the contributions from $\gamma$ and Z bosons,
while $\sigma_L$ is sensitive to $t-$channel sneutrino exchange.
The combination of these measurements allows for a complete exploration
of the parameter space in the chargino sector.

\subsection{mSUGRA}
One of the most widely studied supersymmetric models is the mSUGRA,
or minimal supergravity, model~\cite{msugra}.  This model has the
practical advantage that
it is extremely restrictive, with all masses and mixing parameters being
predicted in terms of four input parameters
 and a sign.  The model assumes that the mass of all of
the new scalar sparticles are given by a single parameter, $m_0$,
at the grand unification scale, $M_{GUT}\sim 2 \times 10^{16}~GeV$ (
$M_{GUT}$ is defined to be the energy scale where the $3$ gauge coupling
constants meet).
The gauginos are likewise assumed to have a common mass, $M_{1/2}$, at
the GUT scale.  At the scale $M_{GUT}$, the theory is completely
specified by,
\begin{equation}
m_{0},~ M_{1/2},~ A_0, ~\tan\beta, ~sign(\mu).
\end{equation}
All of the sparticle masses and mixings can be expressed in terms of
these variables.

Renormalization group equations
are used to evolve all parameters to the weak scale.
At this scale, the gaugino masses are predicted to be in the ratio,
\begin{equation}
M_1:M_2:M_3~=~1:2:7\quad .
\label{mgauge}
\end{equation}
Measurement of gaugino parameters in these ratios would be a strong
indicator for the mSUGRA type models. 
Typically in mSUGRA models, the lightest scalars are the sleptons 
 and over much of the 
parameter space, the lightest supersymmetric particle is the neutralino.

\begin{figure}[ht]
\centerline{
\epsfig{file=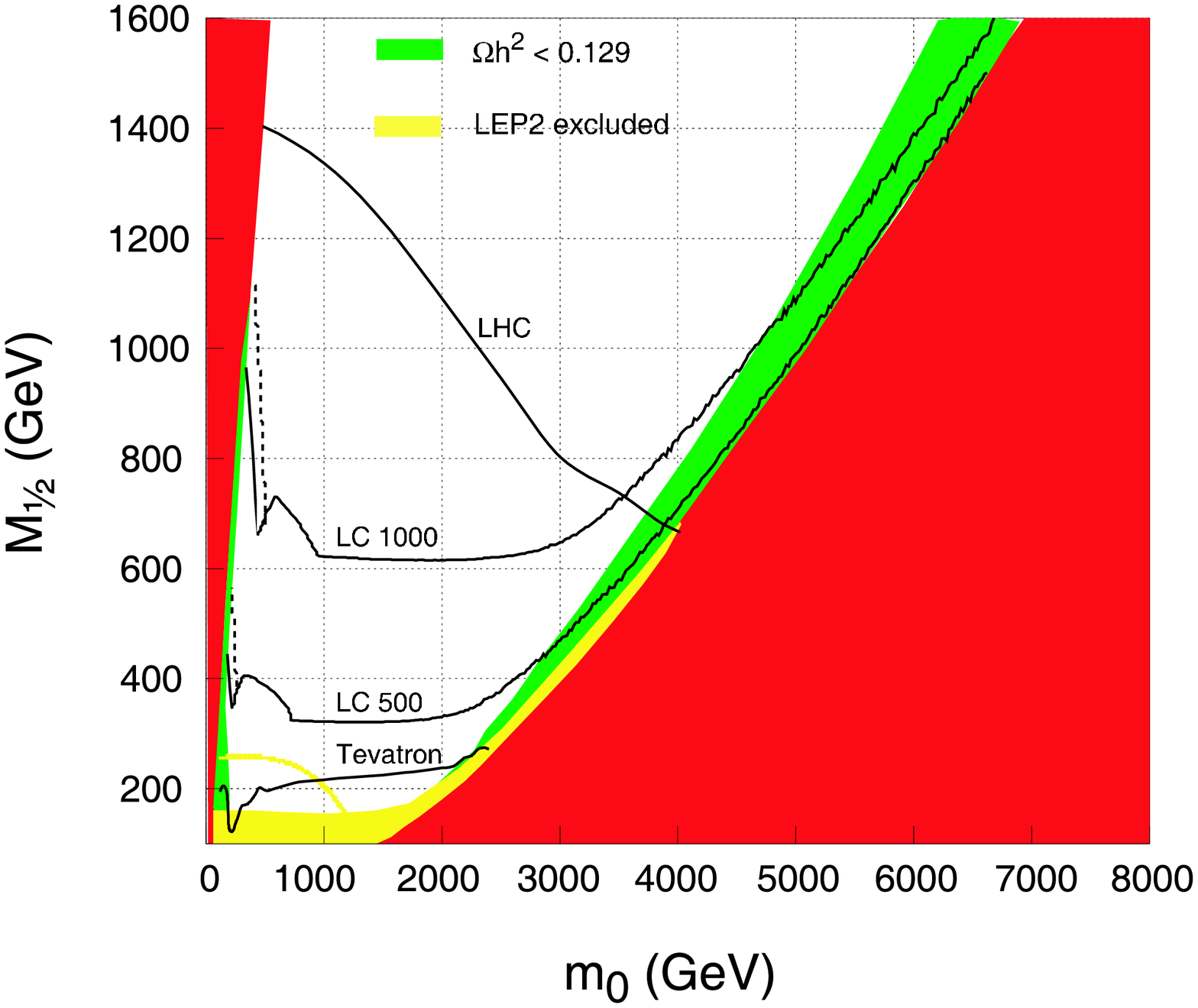,width=4.5in}}
\vspace*{-.1cm}
\caption[]{\it 
Discovery reach of a $\sqrt{s}=$500~GeV and 1~TeV linear
collider for the mSUGRA model with $\tan\beta=30$, $A_0=0$,
and $sign(\mu)=+$.  The discovery reach of the Fermilab
Tevatron with an integrated luminosity of
$10~\mathrm{fb}^{-1}$, and the LHC with $100~\mathrm{fb}^{-1}$ are also shown.
The green shaded region is that preferred by dark 
matter experiments, while the brown region is theoretically
inconsistent.
From reference~ \cite{lcsusy}.} 
\label{fig:reach}
\end{figure}

The discovery reach of a collider in the mSUGRA model can be conveniently 
expressed in terms of its reach in the $m_0~-~M_{1/2}$ 
plane for fixed values of
$A_0$, $\tan\beta$, and $sign(\mu)$~\cite{dd}. 
 An example of such a reach plot
is shown in figure~
 \ref{fig:reach}. It is interesting to note
that  there is a region of parameter space (large $m_0$
and moderate $M_{1/2}$) which is inaccessible to the LHC,
but which a 1~TeV linear collider can explore~\cite{lcsusy}. 
 This region is particularly
interesting because the parameters allow the neutralino to be a
dark matter candidate consistent with astrophysics 
experiments~\cite{susydm}.

\subsection{Gauge Mediated Supersymmetry Breaking}
In this class of supersymmetric models, the supersymmetry breaking
is communicated from 
a hidden sector by 
 the gauge interactions of a new type of particle called a messenger field. 
Gauge mediated models~\cite{gauge_med} have a distinctive pattern of masses:  
gaugino masses arise at one loop and are proportional to the
gauge coupling constants,
\begin{equation}
M_i\sim {\alpha_i\over 4 \pi}{F\over M}
\end{equation}
and scalar masses arise at two-loops
\begin{equation}
M_{scalar}^2\sim \Sigma_i c_i\biggl({\alpha_i\over 4 \pi}\biggr)^2
\biggl({F\over M}\biggr)^2
\end{equation}
where $\sqrt{F}$ 
is the scale of supersymmetry breaking,
$M$ is the messenger mass, and $c_i$ depends
only on the gauge quantum numbers of the
scalar.  This type of model
yields a hierarchy between the strongly interacting squarks and gluinos
and the weakly interaction sparticles.  Confirmation
of this pattern of masses at a linear collider would
be a signal for gauge mediated supersymmetry
breaking.

\subsection{Extracting the Underlying SUSY Parameters}
If evidence for supersymmetry is discovered at either the LHC or at a linear
collider, the next goal will be to understand the source of supersymmetry
breaking.  This symmetry breaking presumably occurs at a high energy
scale. 
As we have seen, the pattern of 
sparticle masses can be quite different depending
on the source of the supersymmetry breaking.
Precise measurements of SUSY
masses and couplings can be used to extrapolate from
the weak scale to high energy using
renormalization group techniques.
Because the
evolution of the masses and couplings with
energy is sensitive to the underlying
model, such an exercize offers a window to the Planck mass scale.

\begin{figure}[ht]
\centerline{
\epsfig{file=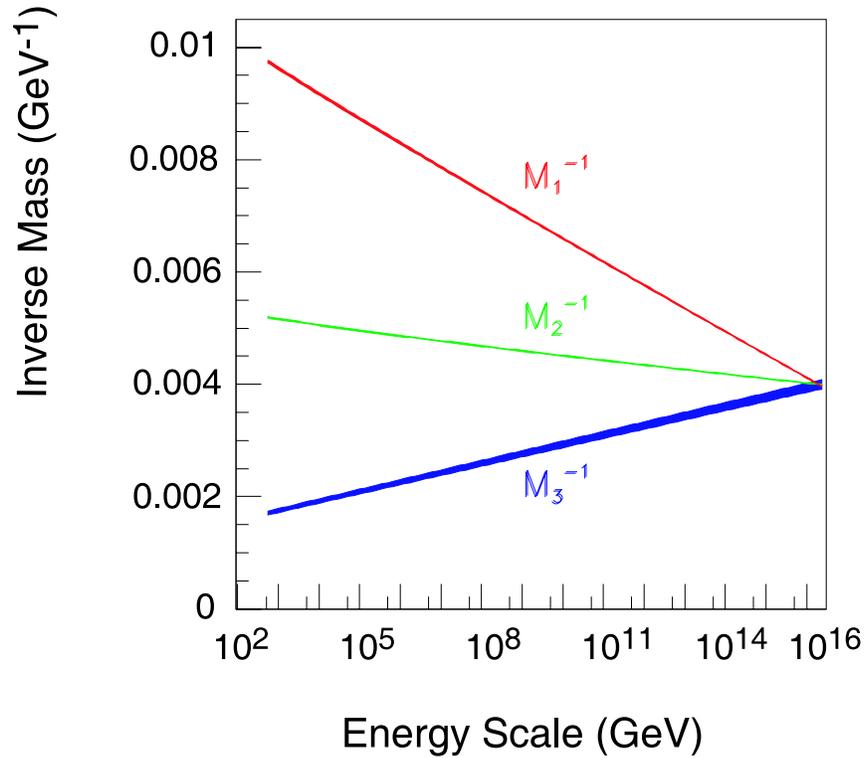,width=4.5in}}
\vspace*{-.1cm}
\caption[]{\it
Evolution of the gaugino masses in an mSUGRA model. The width of the band
is the $1\sigma$ confidence level assuming the experimental accuracies expected
at the LHC for $M_3$ and a linear collider for $M_1$ and  $M_2$. The mSUGRA point
is $m_0=200$~GeV, $m_{1/2}=250~$GeV, $A_0=-100$~GeV, $\tan\beta=10$, and 
$sign(\mu)=+$.
From reference~ \cite{unig}.} 
\label{fig:gaug_evol}
\end{figure}

The evolution of the gaugino masses 
is particularly interesting, since in many models the
ratio of gaugino masses is predicted
to be that of Eq. \ref{mgauge}.  Figure ~ \ref{fig:gaug_evol}
(from \cite{unig})
shows the evolution of gaugino masses in an mSUGRA model, assuming the 
precision which can be obtained in a linear collider
for the measurements of $M_1$ and $M_2$.  Note
that $M_3$, the mass of the gluino, cannot be measured at a linear collider,
but must be determined at the LHC. Similar plots can be made for the
evolution of the scalar masses.   The combination of a linear collider
with the LHC will truly probe the high energy/GUT scale~\cite{gw}.

\clearpage

\section{Other Extensions to the Standard Model}
\label{other}

\subsection{Strongly Interacting Electroweak Symmetry Breaking}

Although the precision measurements shown in figure 1 strongly
indicate the presence of a light Higgs boson, or something
like it, it is possible that the electroweak symmetry breaking
results from a new strong interaction, such as technicolor~\cite{hills},
a composite Higgs theory, or even something
completely different than the Standard Model.  
Theories with a heavy Higgs boson, or
with no Higgs boson, can be 
parameterized in terms of effective
Lagrangians, where the operators ${\cal O}_i$ are constructed from the Standard
Model fields and required to be invariant under the $SU(3)
\times SU(2)\times U(1)$ gauge groups.
The effective Lagrangian describes physics at the weak scale
and the new physics which replaces the light Higgs boson of
the Standard Model is
contained in higher dimension operators
which  then depend on the scale at which the new physics is
introduced, ${\Lambda}$,
\begin{equation}
{\cal L}\sim {\cal L}_{SM}+\Sigma_i {f_i\over \Lambda^2}{\cal O}_i+...
\label{effl}
\end{equation}
Fits to precision electroweak
measurements allow a heavy Higgs boson with
a mass $M_h\sim$~ 400-500~GeV, with new physics at the scale $\Lambda
\sim {\cal O}(1-3)$~TeV~\cite{bfs}. 
Similarly, the Higgs boson can be removed completely
from the theory with new physics occurring
through the operators of Eq. \ref{effl} at the TeV scale~\cite{bfs}.  
In both the case with a heavy Higgs boson and where there is no Higgs boson,
the three-  and four-gauge boson interactions will differ from
those of the Standard Model by terms which scale as 
powers of $1/\Lambda$.

If the Higgs boson is heavy, $M_h\sim$~ 400~GeV or larger,
there is no distinct Higgs resonance, since the
Higgs boson width grows with the Higgs mass.  A heavy Higgs boson
implies that the self-interactions of the W and Z gauge bosons
are becoming strong~\cite{lqt}.
The measurement
of the branching ratios, $\mathrm{h}\rightarrow 
\mathrm{W}^+\mathrm{W}^-$ 
and $\mathrm{h}\rightarrow$~ ZZ, would be
sensitive to this physics and could help to untangle the underlying 
mechanism of symmetry breaking~\cite{barklow}.   
If the Higgs boson is removed completely from the theory, or
is very heavy, then
the effects will be seen in non-Standard Model self- interactions
of the W and Z gauge bosons.  Changes to the 4-gauge boson 
vertices can be probed by measuring gauge boson scattering
through the processes, 
$\mathrm{e}^+\mathrm{e}^-\rightarrow \nu {\overline \nu} W^+ W^-$
and $\mathrm{e}^+\mathrm{e}^-\rightarrow
 \mathrm{e}^+\mathrm{e}^-\mathrm{Z}\mathrm{Z}$.
These effects are suppressed by
powers of the scale $\Lambda$ and are expected to be small~\cite{accomando}.
Probing the 4-gauge boson vertices requires that the linear collider run
at the highest possible energy and with 
the highest possible luminosity.

\begin{figure}[H]
\centerline{
\epsfig{file=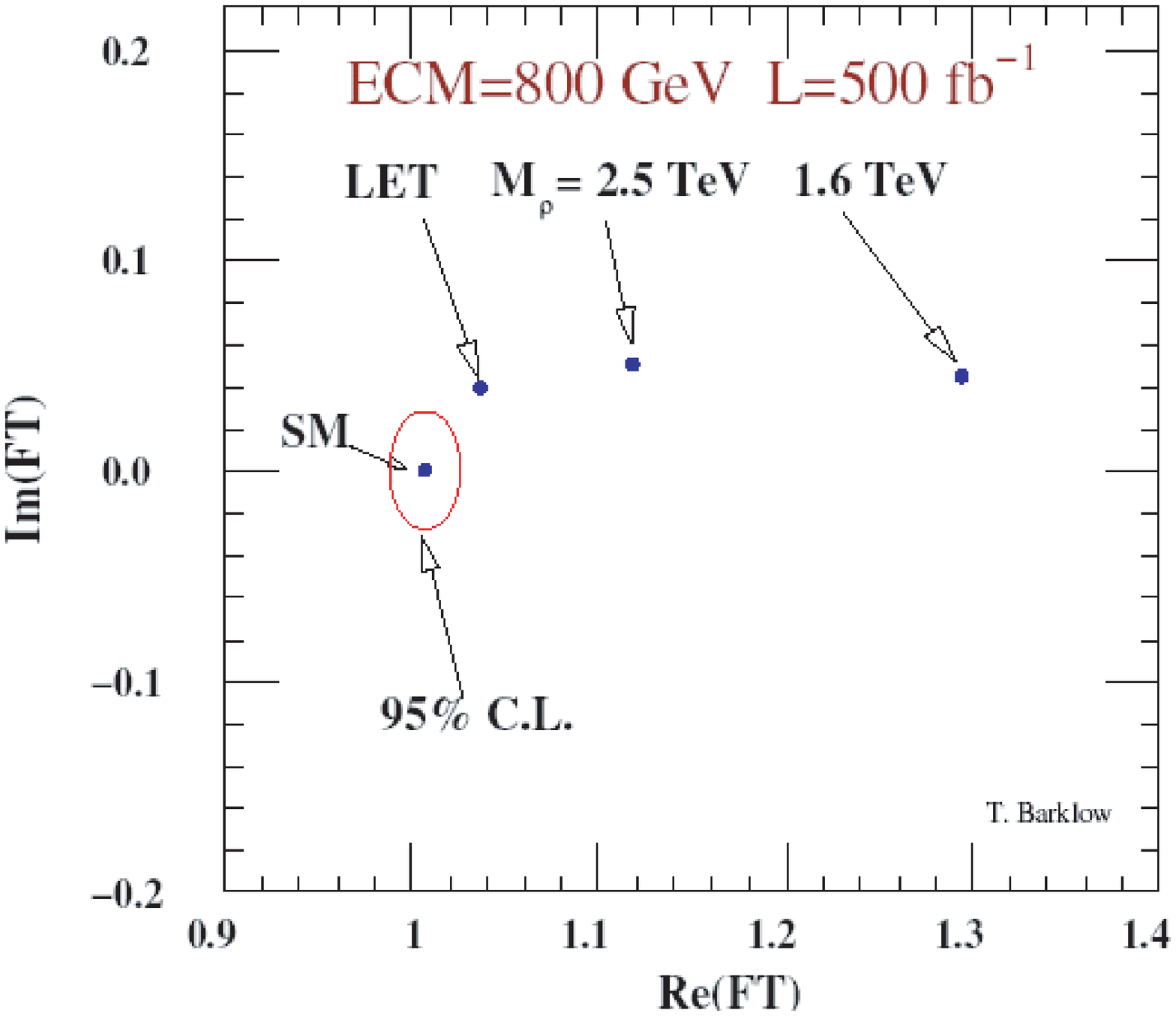,width=4.5in}}
\vspace*{-.1cm}
\caption[]{\it Sensitivity of the 
longitudinal W form factor at a linear collider with
$\sqrt{s}=800~GeV$ and an integrated luminosity of $L=500~
\mathrm{fb}^{-1}$ to
a ``$\rho$''-like resonance of mass $1.6$ and $2.5$~TeV.  Also shown is the 
Standard Model result and the result in a theory where the Higgs mass
is taken to be infinitely massive (labelled LET).
From reference~\cite{accomando}.}
\label{fig:strongfig}
\end{figure}

Strong interactions in the electroweak sector can also manifest
themselves through alterations of the 3-gauge boson vertices.
The process $\mathrm{e}^+
\mathrm{e}^-\rightarrow \mathrm{W}^+\mathrm{W}^-$ is sensitive to alterations
in the $\gamma \mathrm{W}^+\mathrm{W}^-$
 and $\mathrm{Z}\mathrm{W}^+\mathrm{W}^-$ 
vertices.  Since 
 $\mathrm{W}^+\mathrm{W}-$  pair production
 pairs is the largest single contribution to the total
$\mathrm{e}^+\mathrm{e}^-$
 cross section, this process is quite sensitive to small
changes in the 3-gauge boson vertices~\cite{accomando,pm}. In addition, 
in many models with strong electroweak symmetry breaking,
 there is a $\rho$-like resonance with a mass below a few
TeV whose properties can be
modeled by analogy with QCD.  
A linear collider running at $\sqrt{s}=800$~GeV
and with an integrated luminosity of $L=500~\mathrm{fb}^{-1}$ can
see the effects of a 1-2~TeV resonance through the
process $\mathrm{e}^+\mathrm{e}^-\rightarrow 
\mathrm{W}^+\mathrm{W}^-$~\cite{tesla} as is  shown in figure
\ref{fig:strongfig}.\footnote{When the $\mathrm{W}^+
\mathrm{W}^-$ interactions become strong, the amplitude
for $\mathrm{e}^+\mathrm{e}^-\rightarrow 
\mathrm{W}^+\mathrm{W}^-$ develops a complex form
factor, $F_T$. Figure \ref{fig:strongfig} shows the 
predictions for this form factor is a model with a heavy
vectorlike $\rho$ meson.}    A linear collider can even probe the
scenario where the Higgs boson is removed from the theory by
becoming infinitely massive (the point labelled LET 
in figure~\ref{fig:strongfig}).

In order to go beyond the effective Lagrangian approach,
it is necessary to construct
explicit models which allow for a heavy Higgs boson, or no Higgs boson,
and are consistent with the precision electroweak measurements.
These models tend to have  new light vector bosons or fermions whose
effects could be observed at a linear collider~\cite{pw}.

\subsection{Top Mass measurements}

A linear collider, operating near the $\mathrm{t}\mathrm{\bar t}$
 production threshold
energy, will have broad capabilities
for precision top quark measurements. 
Measurements include the top quark mass, $m_t$, 
and width, $\Gamma_t$, along with the top
quark- Higgs boson Yukawa coupling, $g_{tth}$~\cite{fms,mm}. 
The large top quark mass gives 
the top quark a special role in many
models with physics beyond the Standard Model
and so precise measurements of
its  parameters will provide important new insights
into the source of mass~\cite{simmons}.  The top quark mass
is also an important component of the electroweak precision
measurement program, as seen in figure 1, and so it is
important to measure it as accurately as possible.

\begin{figure}[HTBP]
\centerline{
\epsfig{file=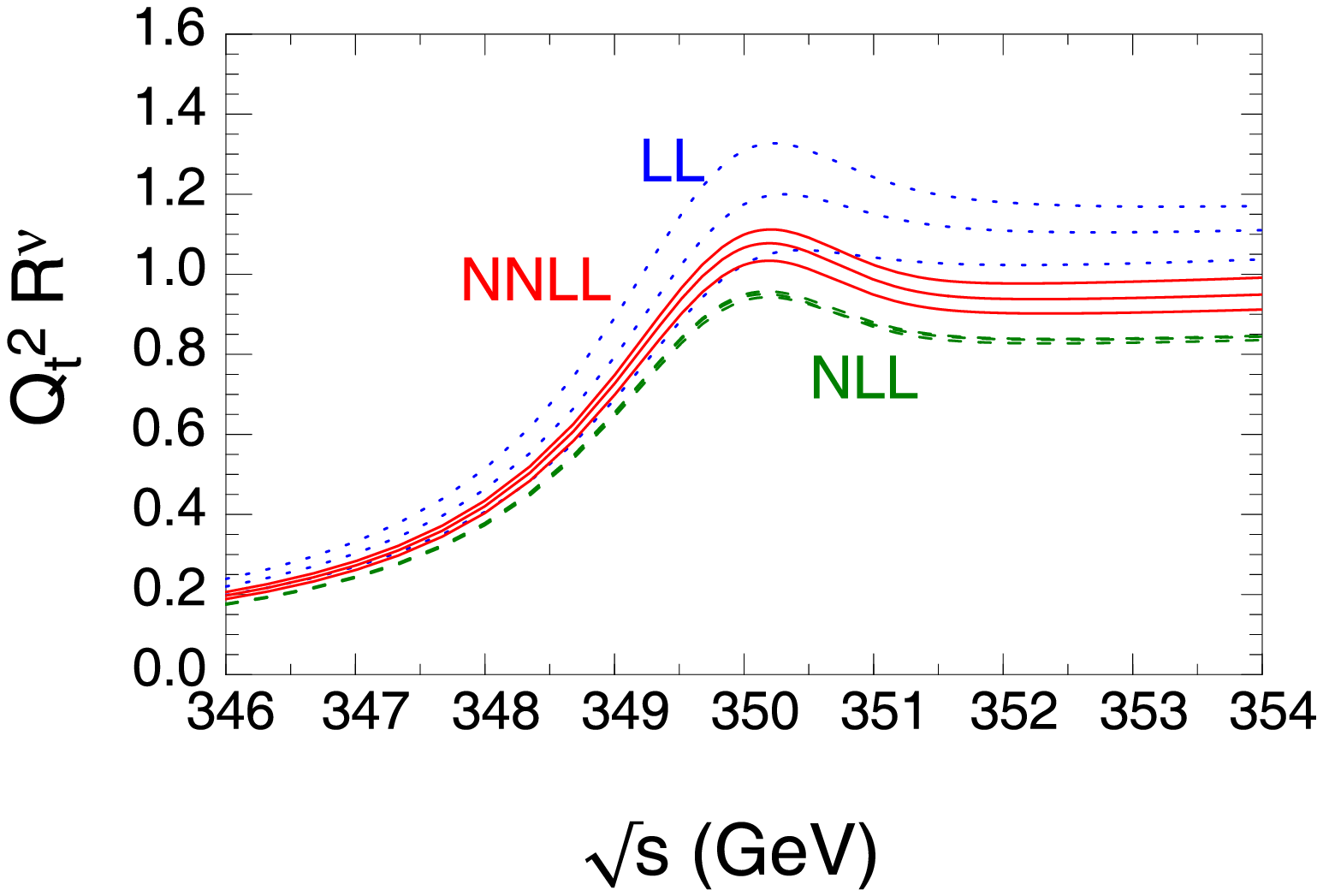,width=4.5in}}
\caption[]{\it 
Dependence of the $\mathrm{e}^+
\mathrm{e}^-\rightarrow \mathrm{t}
\mathrm{ {\overline t}}$ cross section
on the center-of-mass energy, $\sqrt{s}$.  The curves labelled LL
(dotted), 
NLL (dashed), and NNLL (solid) include leading logarithmic, next-to-leading
logarithmic, and next-to-next-to- leading logarithmic contributions,
respectively.  The three curves for each set of corrections represent
the variation of the rate with  changes of the renormaliztion
scale.   This figure uses a threshold mass definition for
the top quark mass~\cite{tmass}. From reference~\cite{hoang}.}
\label{fig:topfig}
\end{figure}

Because of its large width, $\Gamma_t\sim 1.5$~GeV, the top
quark will decay before it hadronizes and so non-perturbative
effects are expected to be highly suppressed. The dependence
of the cross section for $\mathrm{e}^+
\mathrm{e}^- \to \mathrm{t} \mathrm{\bar t}$ on the energy 
can be computed reliably and the
rate increases by a factor of ten
as the center- of- mass energy is varied by 5~ GeV 
around the threshold energy.  The location of the rise of 
the cross section can be used to extract the top quark mass, while
the shape and normalization yield information
about $\Gamma_t$, $g_{tth}$, and $\alpha_s$.
The threshold cross section has been calculated 
including some of the next-to-next-to-leading 
logarithms, 
as shown in figure \ref{fig:topfig}~\cite{hoang,tmass}. The complete set
of next-to-next-to-leading logarithmic contributions is not
yet complete, but the large size of the corrections
relative to the next-to-leading logarithmic terms~\cite{hoang2}
 suggests
that the uncertainty on the cross section measurement will
be slightly larger than previously estimated,
$\delta\sigma_{tt}/\sigma_{tt}\sim \pm 6\%$.  This implies that
a linear collider operating at the 
$\mathrm{t} \mathrm{{\overline t}}$ threshold will
be able to measure the top quark mass with an accuracy of $\delta
m_t\sim 100$~MeV~\cite{mm}.  This should be compared with an expected
accuracy at the LHC of $\delta m_t\sim$~ 1-2~GeV~\cite{atlastdr}.



\subsection{Extra Dimensions}

An alternative solution to the hierarchy problem was discovered
in 1998~\cite{ADD} through the realization that an effective Planck
%
%
scale with electroweak energy can exist as a fraction of the true Planck
energy if there are extra spatial dimensions.
There are now three classes of these extra-dimension models,
differing from each other in the geometry of the various fields.
They also have different experimental signatures~\cite{0007226}.
%
%
Common features of the models are closely-spaced excitations extending
into the TeV mass range 
and particles with spin-2, 
both well suited for study in $\mathrm{e}^{+}\mathrm{e}^{-}$
collisions at the highest possible energies.
Furthermore, the $\mathrm{e}^{-}
\mathrm{e}^{-}$, $\mathrm{e}^{-} \gamma$, and $\gamma \gamma$
options of a linear collider~\cite{optns},
%
%
as well as the possibility of transverse polarization~\cite{0211374},
are uniquely suited to differentiating
the extra-dimension models.
%
%

\begin{figure}[htbp]
\epsfig{file=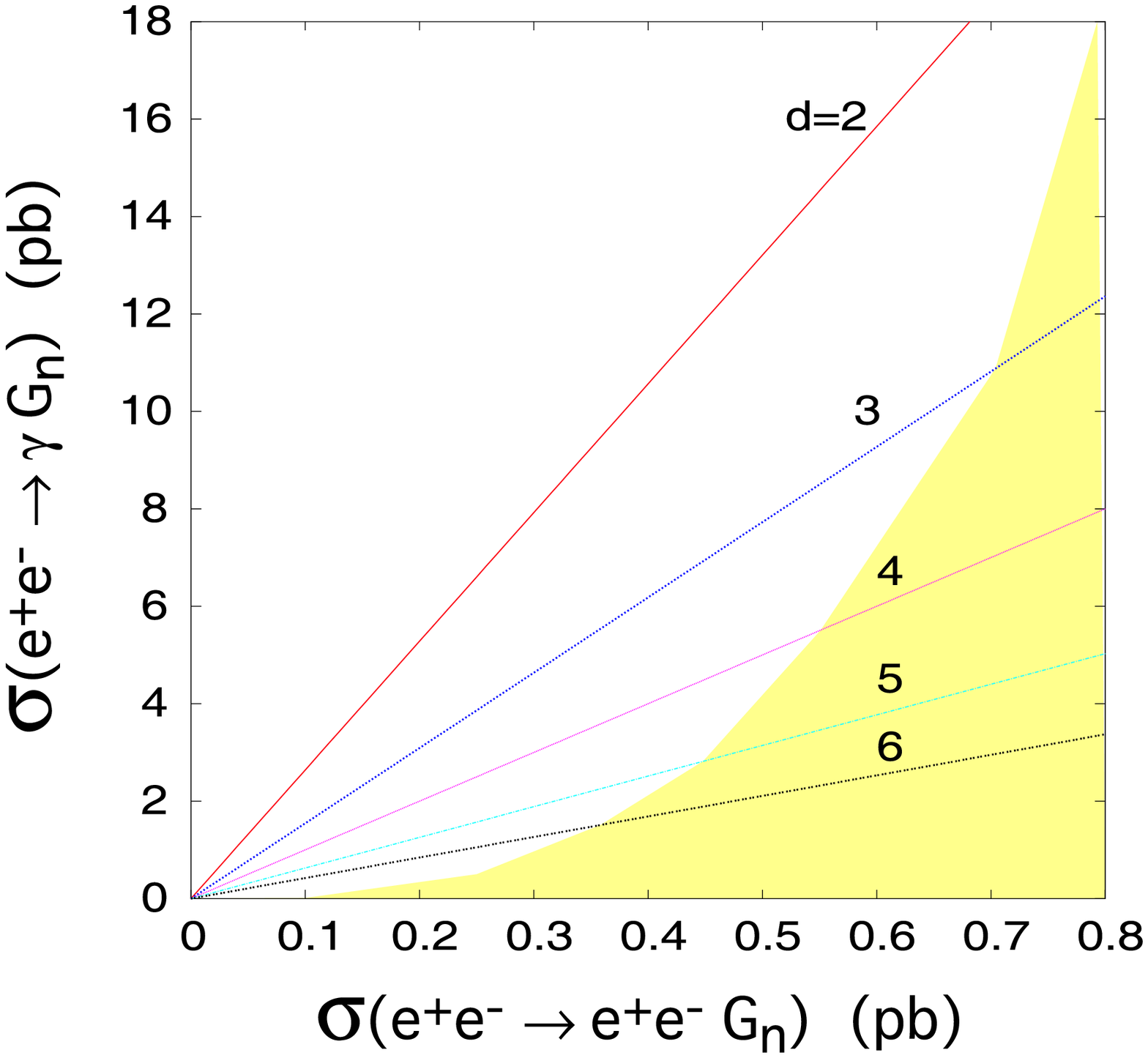,width=4.5in}
\caption
{\it
Cross section for
$\mathrm{e}^{+}\mathrm{e}^{-} \rightarrow \gamma G_{n}$
versus
$\mathrm{e}^{+}
\mathrm{e}^{-} \rightarrow \mathrm{e}^{+}\mathrm{e}^{-} G_{n}$
for polarized electrons
at $\sqrt{s} = 500$~GeV;
$G_{n}$ are the KK excitations of the gravitons.
The number of extra dimensions is indicated by the number on each curve.
The shading represents the region of low reliability
for the theoretical calculation.
From reference~\cite{0307117}.
}
\label{fig:bbg}
\end{figure}

The original extra dimensions model was described by
Arkani-Hamed, Dimopoulos, and Dvali (ADD)~\cite{ADD}; 
it confines the Standard Model fields to a 1+3
dimensional brane, while gravity propagates in the ``bulk'' of the
other dimensions. In this way, gravity might be strong in the bulk,
but only a fraction is felt on the brane.  
It can be arranged so that d of the dimensions forming the bulk 
share a relativly large length scale R, with the result:
\begin{equation}
M^{2}_{planck} \sim M^{d+2}_{brane} R^{d},
\label{mplanck}
\end{equation}
where $M_{brane}$ is the effective Planck mass scale on the 3-brane
where the Standard Model lives.
Thus, the effective Planck mass can be tuned  TeV scale;
for d=2, R is on the order of a millimeter.
Unfortunately, ADD introduces a new hierarchy problem 
insofar as the TeV string scale is so different from the
compactification scale.
A consequence of the graviton propagating in the bulk
is a tower of excited Kaluza-Klein (KK) graviton states
which couple weakly to Standard Model particles.
In $\mathrm{e}^{+}\mathrm{e}^{-}$ collisions 
the experimental signature is either missing
energy in Standard Model processes by radiation of a KK graviton 
from SM particles~\cite{0103053}, 
%
%
or altered pair production by virtue of a graviton
propagator~\cite{9811291,9811350,0110346,9902263,0010354}
(the pair rate and angular distribution would be affected).
%
%
%
%
%
%
A striking signature would be radiation of a graviton from a virtual
photon, creating a visible photon and missing energy~\cite{9811337,0110339}.
%
%
%
Figure~\ref{fig:bbg} shows a calculation~\cite{0307117}
for the radiated graviton in Bhabha scattering.
Clean measurement of the angular distribution of final states
consisting of a single photon and missing energy should be able to
distinguish the ADD model from 
a scenario where the missing energy
is from extra neutrinos or superpartners.

%
%

\begin{figure}[htbp]
\epsfig{file=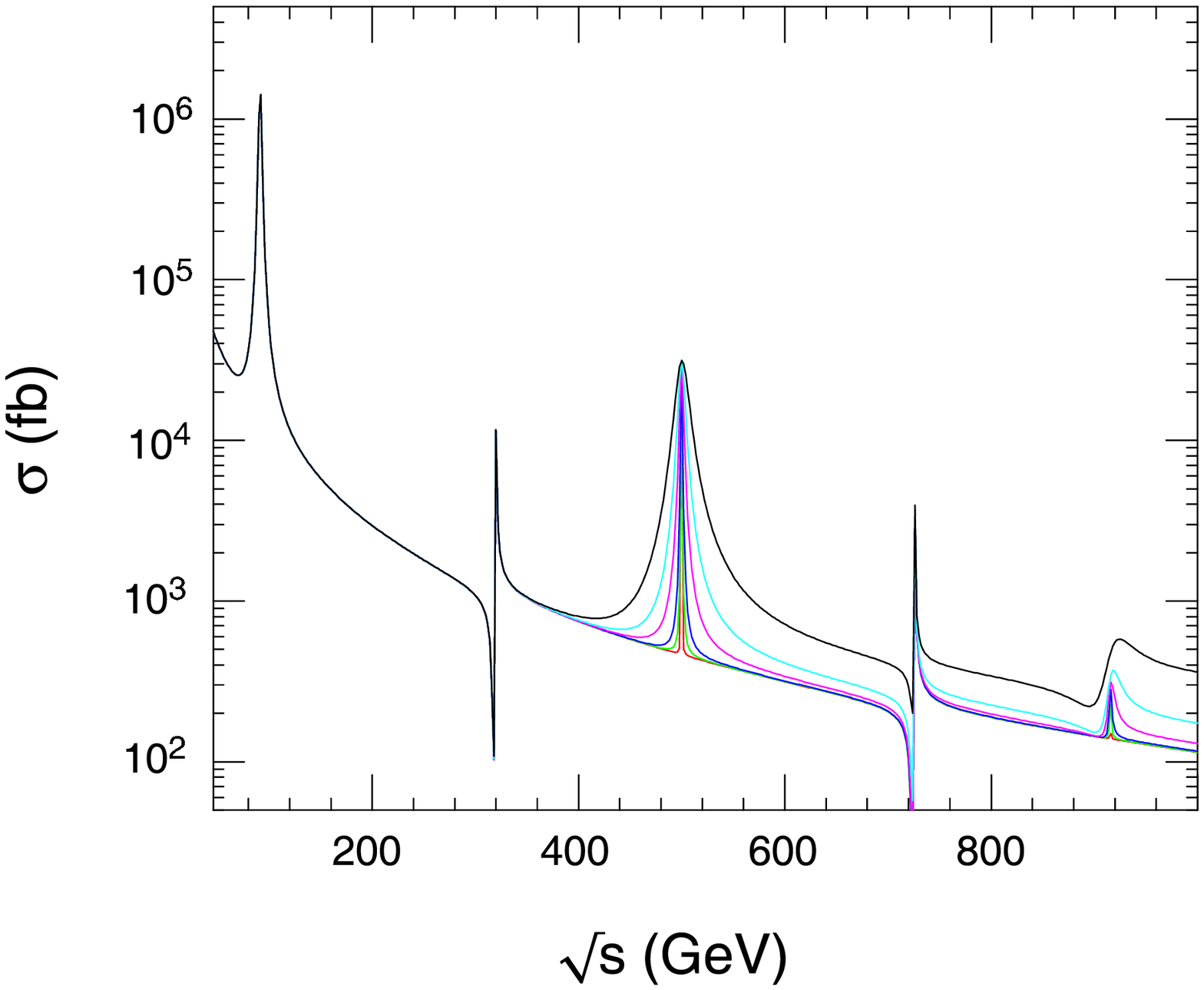,width=4.5in}
\caption{
$\mathrm{e}^{+}\mathrm{e}^{-}$ production of graviton and neutral gauge KK excitations
for a model having the first KK excitation at 500~GeV.
The series of curves represents various values of $\kappa/\bar{M}_{planck}$.
From reference~\cite{0006041}.
}
\label{fig:hewett}
\end{figure}

\begin{figure}[htbp]
\epsfig{file=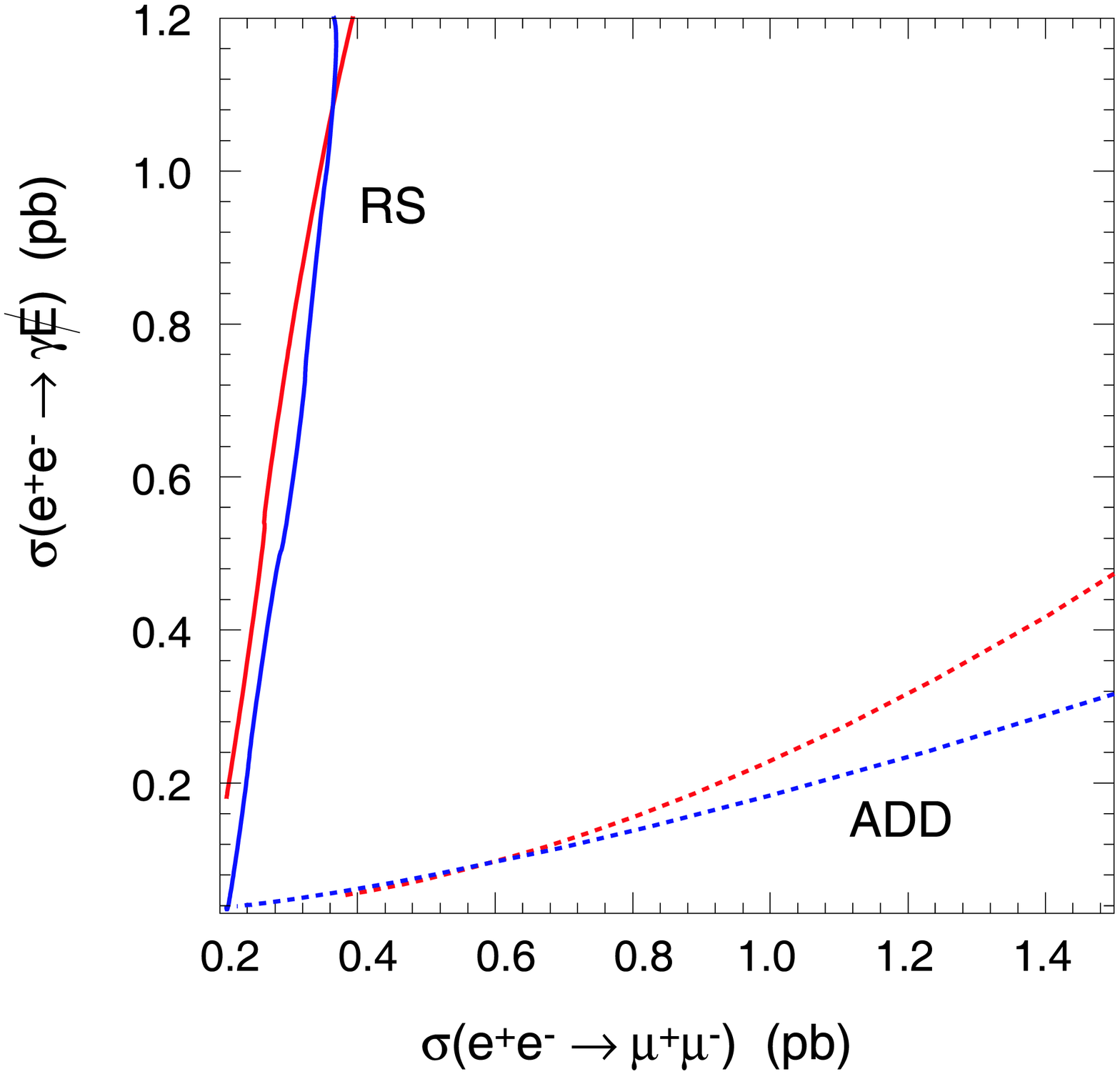,width=4.5in}
\caption{\it
Cross section for single photon production compared to the dimuon
production for some ADD and RS models
in $\mathrm{e}^{+}\mathrm{e}^{-}$ production at $\sqrt{s}$ = 2~TeV.
The two ADD lines are for extra dimensions of 3 and 6.
The RS curves are for $m_{0}$ = 200 and 400~GeV.
From reference~\cite{0307096}.
}
\label{fig:sr}
\end{figure}

A different model 
due to Randall and Sundrum (RS)~\cite{rs}
allows for a brane having a TeV scale
without requiring the compactification radius of the extra dimensions
to be large and cures the ADD heirarchy problem. 
In the RS schemes
%
%
there is one extra dimension and there are two branes, one with 
an effective Planck mass ($\bar{M}_{planck}$) which can be the TeV scale, 
and the other at the Planck scale; a scalar field in the bulk
is needed to keep the branes the proper distance apart.
This results in a factor 
$e^{- \pi \kappa R }$ which warps the spacetime metric and creates the
scale difference between the two branes; 
$\kappa$ is the curvature of the fifth dimension and R is the
compatification scale.
Since the metric is a function of the 5th coordinate, the theory is nonfactorizable.
A TeV scale can be achieved on the low energy brane if 
$\kappa R \sim 12$.
The original
models had the Standard Model  particles located on the TeV brane,
 but current models
let all particles apart from the
Higgs boson reside in the bulk.
As in the ADD case, the extra dimension gives rise to KK excitations
of the Standad Model particles.
Unlike ADD, the graviton masses are of the electroweak
scale, and there will be a series of KK gravitons.
The masses of the KK excitations of the gravitons are
$M_{n} = \zeta_{n} m_{0}$, where
$m_{0} = \kappa e^{- \pi \kappa R}$ and $\zeta_{n}$ are zeroes of the
Bessel function~\cite{0006041}; other schemes lead to different mass spacings.
%
%
It is perhaps more useful to consider the free model parameters as
$m_{0}$ and $c = \kappa / \bar{M}_{planck}$ which is an effective
coupling~\cite{0307096}.
%
%
The gravitons have weak strength and
decay into pairs of Standard Model
particles  with a width dependent on $c_{0}$.
Figure~\ref{fig:hewett} shows a spectrum of such KK excitations,
clearly well suited to the excellent mass resolution of a linear collider.
These KK resonances resemble excited Z bosons. 
The key RS experimental signature is the spin-2 nature of the RS
gravitons, which is reflected in the angular distributions
and often with an enhancement of particles transverse to the beams.
Figure~\ref{fig:sr} summarizes the difference between ADD and RS
production of single photons versus dimuons. At the very high energy
scale of $\sqrt{s}=$2~TeV shown in this figure, it is possible for a linear
collider to distinguish between these two extra dimension scenarios. 

The RS bulk scalar field gives rise to a projection on the TeV brane;
the resulting particle is dubbed the ``radion''~\cite{rad}.
%
%
The radion manifestation of the scalar field should also have clear
experimental consequences due to its Higgs-like nature. 
The radion field therefore mixes with the Higgs boson
and can significantly change the Standard
Model Higgs boson branching ratios.

There is also the possibility that all of the 
Standard Model fields live in the
extra-dimensional bulk, in which case all 
Standard Model particles will have KK excitations; 
such models are called Universal Extra
Dimensions (UED)~\cite{0012100}.
%
%
%
In UED models, because all the fields are allowed in the bulk,
conservation laws apply with the consequence that there is KK number
conservation. At lowest order, excited particles are produced in
pairs,
and the lightest KK particle is the photon with n=0.
With excitations possible for all particles, the phenomenological
space is large and more study is encouraged.
This model is actually very similar to SUSY insofar as every SM
particle has a tower of KK-partners; in contrast to SUSY, the
partners have the same spin. Determining whether one is seeing UED or
SUSY will require excellent determination of angular distributions.
In the UED models there can be excited quark bound states 
resulting in a complex spectrum~\cite{0312055}.
%
%

%
%
%

\clearpage

\section{The Linear Collider and Cosmology}
\label{astro}

\subsection{The Linear Collider and Astrophysics}

\begin{figure}[htbp]
\epsfig{file=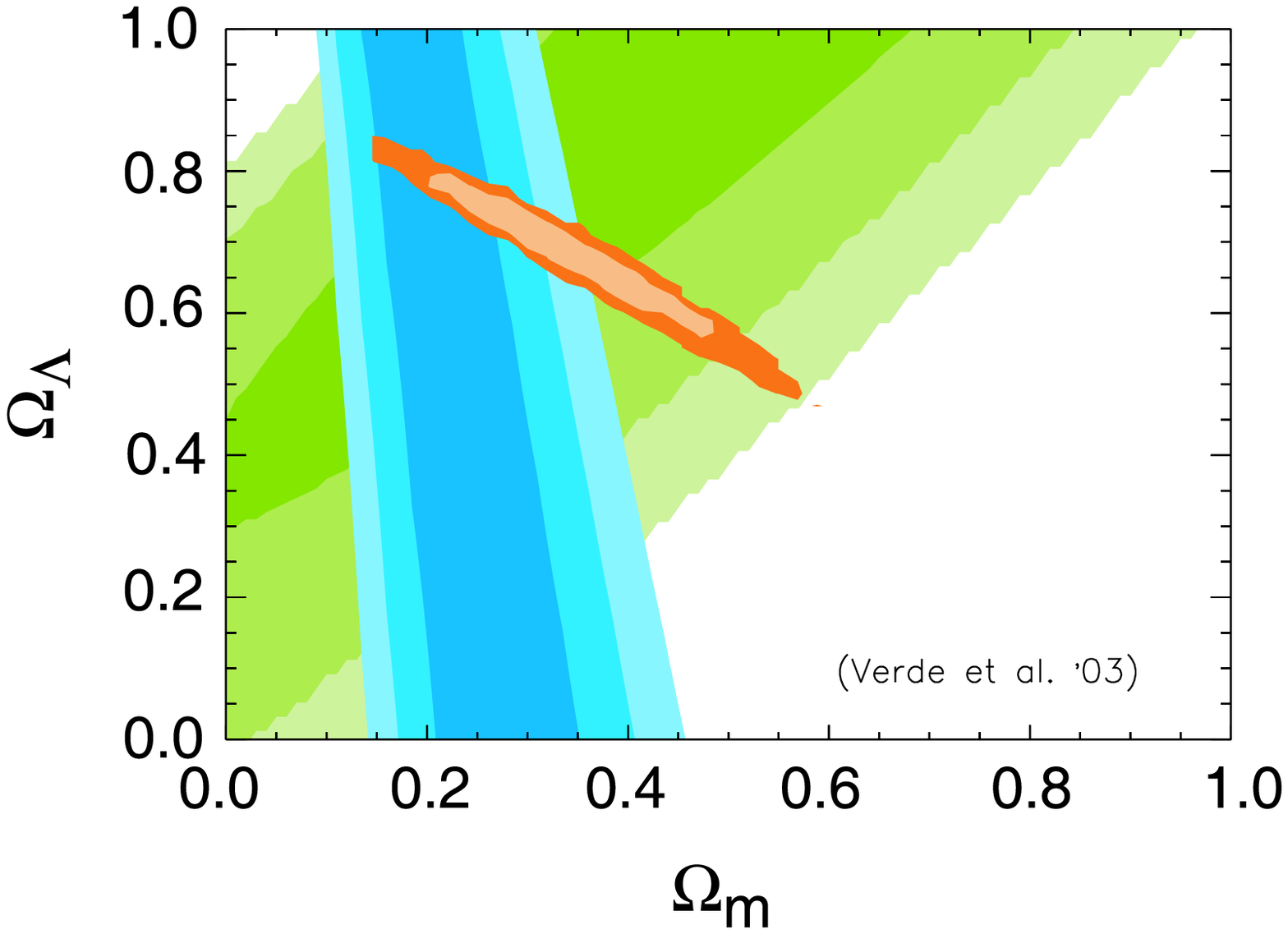,width=4.5in}
\caption
{\it
The one and two standard deviation contours
of experimental data
in $\Omega_{\Lambda}-\Omega_{m}$ space.
The broad band with positive slope represents supernova data.
The narrow band with negative slope comes from CMB anisotropy
measurements.
The vertical band is from the $2^{\circ}$ field galaxy survey.
From reference~\cite{verde}.
}
\label{fig:verde}
\end{figure}

Two startling discoveries in astrophysics have presented 
High Energy Physics with new challenges.
Astrophysical measurements indicate that 95\% of the gravitating energy in the universe
is of an unknown form~\cite{astrorev}.
%
%
Even more embarassing is the indication that $\sim$70\% of the closure
density of the universe is a dark energy heretofore unpredicted in any
physical theory (except for Einstein's cosmological constant)~\cite{lindner}.  
These conclusions result from a number of remarkably consistent astrophysical measurements. 
Measurement of the Cosmic Microwave Background (CMB)~\cite{HuURL} 
%
%
gives the photon density of the universe.
Together with a value for the Hubble expansion parameter,
measurement of the CMB temperature anisotropy directly measures the
total matter+energy density $\Omega_{tot}$
(scaled by the critical density) 
as well as the density for baryonic matter $\Omega_{B}$.
The baryonic matter is that from all strongly
interacting particles; their density (relative to the photons) is also
inferred from measured isotopic abundances interpreted via big-bang
nucleosynthesis. 
The total matter density can be inferred from 
the speed distribution of luminous matter within galactic clusters,
gravitational lensing,
and features in the CMB anisotropy.
The dark energy is indicated by nonlinearity in
the luminosity-distance curve measured for Type Ia
supernovae~\cite{perlmutter3}.
%
%
Combining the experimental data
yields the values 
0.044$\pm$0.004
for the baryonic matter density, 
0.22$\pm$0.02
for the dark matter density, and
0.73$\pm$0.04 for the dark energy 
component~\cite{bennett,spergel,eoss,ccn}.
%
%
%
%
%
The remaining 0.006 constitutes neutrinos and photons.
One particular example of the parameter space
is illustrated in figure~\ref{fig:verde} (from reference~\cite{verde}).

%
%

The connections to particle physics are obvious.  
Candidates for the dark matter exist in a number of the extensions to
the Standard Model.  
The dark energy issue is tougher to
understand, but it could possibly be explained
by evolution of the vacuum.
The connection between High Energy Physics and cosmology is not limited to
dark matter and  dark energy.
The Standard Model in its minimal form fails to explain the lack of
observed antimatter in the universe. It is possible that this
asymmetry is created at the weak scale and so linear collider physics could shed
light on the problem.
The spectrum of cosmic rays at ultra-high energy ($10^{19}$~eV and higher)
is not fundamentally understood and could have its roots in electroweak
physics.
In the following sections, we discuss the implications for LC physics
of the astrophysics connections.

\subsection{Dark Matter}
 
Candidates for dark matter~\cite{jgk,fn} 
%
%
abound in the new theoretical extensions to
the Standard Model,
and the role of the LC would be to identify the creation of such
particles and show that the annihilation cross sections, $\sigma_A$,
 are consistent
with the astrophysical measurements.
The only thing we know about the dark matter is 
that it must be neutral and its abundance,
which can be related to its production cross section.
Within the context of cosmological evolution (the big bang coupled
with an electroweak phase transition leading to inflation),
particles are in equilibrium as long as their interaction rate 
with the cosmic plasma is
larger than the Hubble expansion.
When the interaction rate falls below the expansion rate of the
universe, these particles are said to ``freeze out'', implying
that their comoving number density becomes reasonably constant.
For relativistic particles of mass m, 
the number density after freeze-out will be similar to that for the
photons, with 
$\Omega h^{2} \sim m/(100~eV)$,
where h is the Hubble constant in units of 100~km/s/Mpc.
This cannot account for the measured dark matter abundance,
so we must consider nonrelativistic particle freeze-out.
For particles which are nonrelativistic at the time of decoupling,
iterative solution of the Boltzmann equation yields
$\Omega \sim 400~fb/\sigma_A$~\cite{feng0012277,astrorev}.
%
%
Equating this to the measured 0.22 indicates that the dark matter is
nonrelativistic, or cold (CDM), 
and has an annihilation cross section in the weak interaction range.

%
%
There is a growing list of candidates to account for the CDM: 
weakly interacting massive particles (WIMPs),
the gravitino,
axions, 
super-weakly interacting particles,
other non-WIMP matter,
and 
excitations arising in extra-dimension models.
The axion is a long-lived pseudo-Nambu-Goldstone boson invented as a possible
solution of the strong CP problem~\cite{strongCP}.
%
%
Axions have very small couplings to matter and therefore
cannot be effectively probed at particle colliders.
The lightest particles in Universal Extra Dimension
theories (Kaluza-Klein towers or brane excitations) are newer
candidates for the CDM~\cite{0206071,0302041,CFM}.
%
%
If these candidates have masses in the electroweak mass
range, the superior mass
resolution of a linear collider would identify the series of narrow
states associated with these dimensional excitations.
WIMPs
are the dark matter candidates for which the LC role
has been most studied so far.
WIMP CDM candidates include supersymmetric particles~\cite{EFGO},
with
the best candidate for CDM being the 
lightest supersymmetric particle (LSP),
which is stable owing to R-parity conservation.
If SUSY breaking is gauge-mediated, the LSP is the gravitino,
which would interact at gravitational strength. This scenario
would be born out experimentally by identifying the other
superparticle masses and couplings.

To demonstrate that an LSP constitutes the CDM, the correct freeze-out
density must be achieved through the possible annihilation and
production scattering processes.
Neutralino annihilation occurs both through self-annihilation and by
scattering with other superpartners (coannihilation) such as staus.
The supersymmetric
 particle spectra can complicate the freeze-out scenario
significantly, therefore
associating the LSP with the CDM will require detailed knowledge of the
supersymmetric particle masses and couplings.
For instance, if the LSP has a slightly heavier partner with the same
quantum numbers and a larger annihilation cross section, the effective
LSP annihilation rate is significantly altered. Simply knowing the LSP mass
and self annihilation cross section is not good enough to identify
this
as the CDM, and this stresses the need for precision measurements in
this physics.
The neutralinos are ideal objects to study at the linear collider.
The masses will be measured to high precision via the kinematics,
and measurement of the production rates from polarized beams will
absolutely establish their gaugino properties. High precision
measurements afforded by a linear collider can determine the effective
$\Omega_{DM}$ to an accuracy nearly equalling that for the astrophysical
measurements~\cite{birkedal}.
%
%

\begin{figure}[htbp]
\epsfig{file=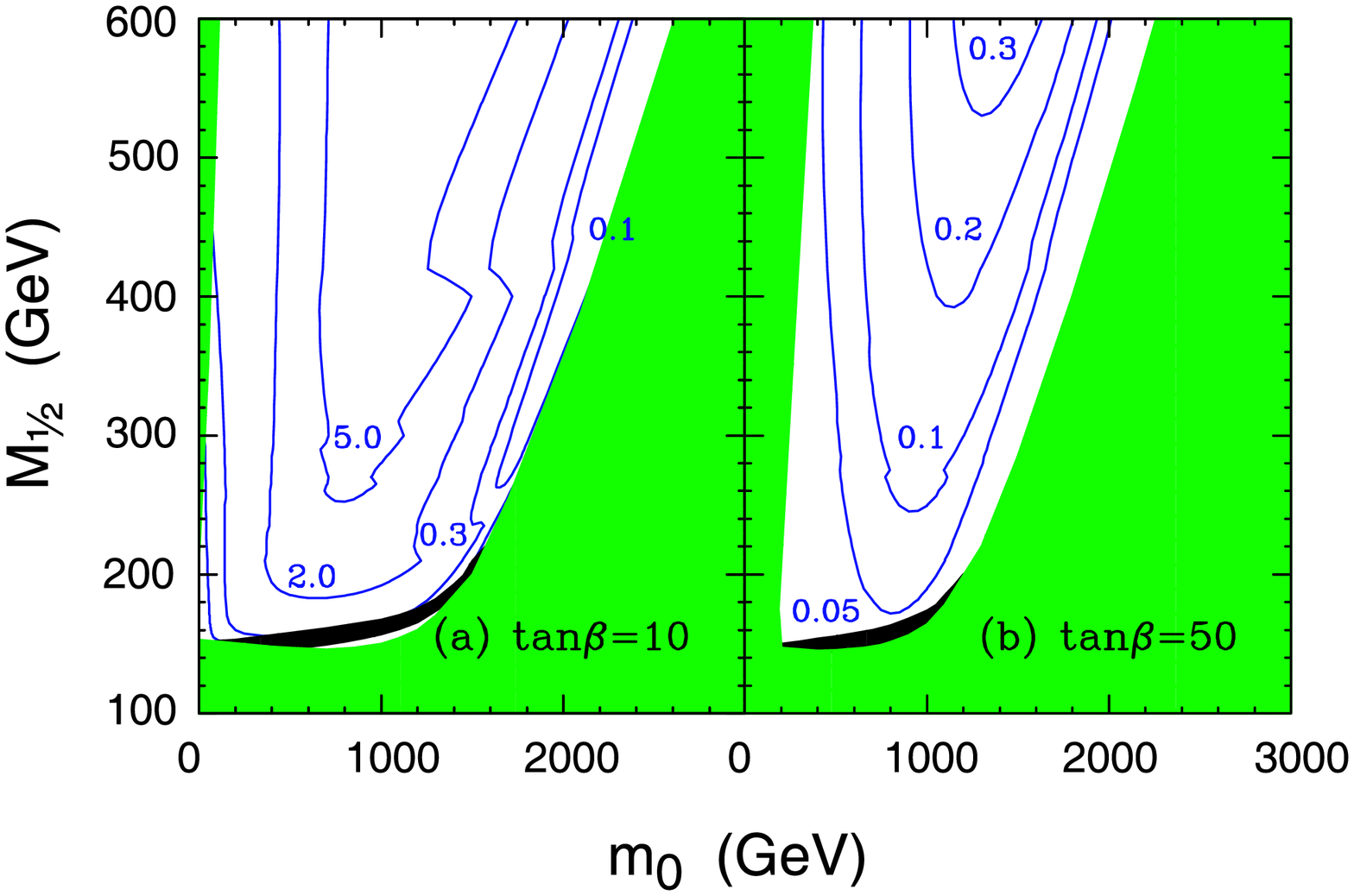,width=4.5in}
\caption
{\it Contours of $\Omega h^2$ in the mSUGRA model.
Here $A_0=0$ and $\mu>0$; representative values of $\tan{\beta}$ are as
indicated.  The shaded regions are excluded by the requirements of a
neutral LSP (left) and the 103 GeV chargino mass bound
from direct searches at the Tevatron (right and
bottom). In the black region, neutralinos annihilate through the light
Higgs pole. (Heavy Higgs poles also play a role in limited regions
with $\tan{\beta}=50$ and $m_0 < $1~TeV.)  Effects of co-annihilation,
important along the boundaries of the excluded regions, have not been
included. 
From reference~\cite{feng0012277}. }
\label{fig:omegaHsq}
\end{figure}

As an example of how the supersymmetric parameter space can be studied
in the context of CDM, we focus on the 
supergravity models (see section~\ref{SUSY}.3). 
In minimal supergravity (mSUGRA) one assumes a single common 
gaugino mass $\mathrm{M}_{\frac{1}{2}}$ and a universal trilinear term $A_{0}$.
The only remaining parameters of this model are 
the soft scalar mass $m_{0}$,
the ratio of Higgs field vacuum expectation values $\tan{\beta}$, and 
the sign of the Higgs mixing mass $\mu$. 
For assumed values of $\tan{\beta}$, $A_{0}$, and sign($\mu$),
the LSP production can be calculated in the 
$\mathrm{m}_{0}$-$\mathrm{M}_{\frac{1}{2}}$ space
to derive values of  $\Omega_{DM}h^{2}$;
this is shown in
figure~\ref{fig:omegaHsq} 
for $A_{0}$=0 and $\mu > 0$.

\begin{figure}[htbp]
\epsfig{file=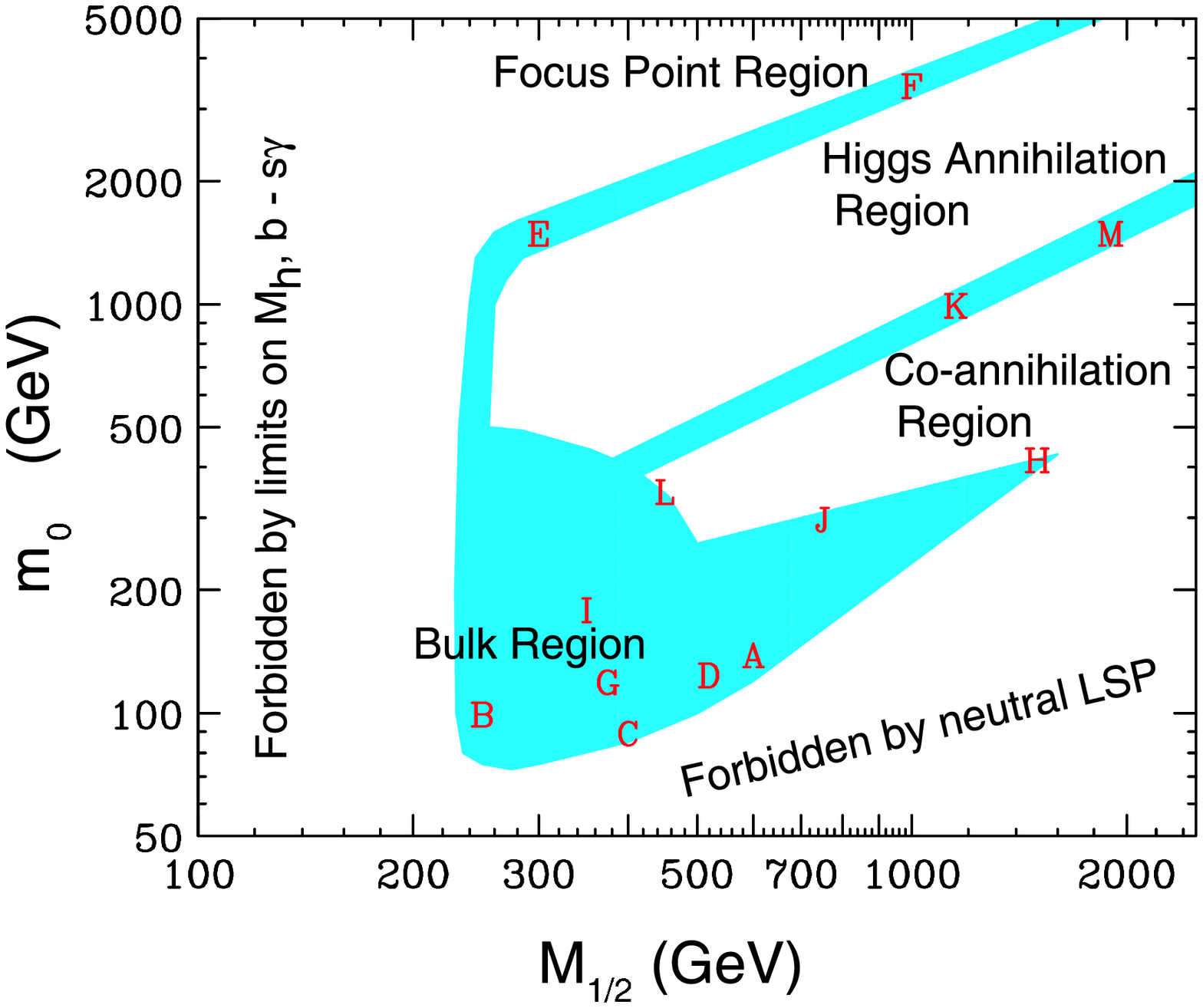,width=4.5in}
\caption
{\it Allowed regions for $\mathrm{m}_{0}$ versus $\mathrm{M}_{\frac{1}{2}}$
in mSUGRA models.
The various labelled regions are described in the text.
Also shown are benchmark points 
for studying various types of supersymmetric model. 
Adapted from references~\cite{fr,lcsusy,ellis0210052}.
  }
\label{fig:benchMJO}
\end{figure}

To explore the mSUGRA possibilities, it is useful to describe regions
of distinct LSP production and annihilation
which can support the observed CDM equilibrium density. 
Figure~\ref{fig:benchMJO} 
(adapted from \cite{fr,lcsusy,ellis0210052}),
%
%
%
shows the 
allowed $\mathrm{m}_{0}$-$\mathrm{M}_{\frac{1}{2}}$ space
after applying restrictions for the feasibility of 
electroweak spontaneous symmetry breaking and 
CDM constraints.
There are four distinct LSP production regions.
In the low mass bulk area the  parameters
of the supersymmetric model are moderate
and in a range where the symmetry breaking does not result from
fine-tuned cancellations between very large sparticle masses.
In this region the LSP is the 
fermion partner of the U(1) gauge boson (the Bino), 
and the model has light sleptons
which allow t-channel annihilation to keep $\Omega_{DM}$ in the
allowed range.
The ``co-annihilation'' region is where the LSP is near-degenerate
with other sparticles, enhancing the annihilation reaction.
The ``focus point'' region at $m_{0} >> \mathrm{M}_{\frac{1}{2}}$
has the 
fermion partner of the neutral Higgs boson (Higgsino) 
as the LSP; here annihilation is enhanced by Z
exchange.
The ``Higgs annihilation'' region is where s-channel annihilation
becomes enhanced for $M_h \sim 2 \mathrm{M}_{LSP}$.
These regions are indicated in the figure, along with letters
indicating suggested benchmark points for studing the possible 
scenarios~\cite{ellis0210052}.

\subsection{Baryogenesis}

Standard Model physics cannot explain the observed matter/antimatter
asymmetry of the universe.  
The theoretical recipe for generating a
baryon asymmetry was pointed out by Sakharov~\cite{sakarov} as
%
%
requiring violation of baryon number,
violation of the C and CP symmetries, and
a departure from thermal equilibrium.
These conditions are possible in GUT-scale
processes~\cite{GUTBAU}, but physics at this scale is not testable in
%
%
particle accelerators.
However, since there is already speculation that supersymmetry
 is broken at the
electroweak scale, electroweak physics does provide several possibilities for
baryogenesis which a linear collider could probe.

In electroweak theory baryon number can be violated by nonperturbative
processes arising from the nontrivial structure of the electroweak
vacuum~\cite{0302151}.
%
%
A phase transition in the Higgs field could give rise to baryogenesis
at TeV temperatures.
The order of the transition is determined by the modulus of the Higgs
field  and the critical temperature is that where the nonzero minimum
of the Higgs field becomes the favored vacuum state. Thermal
equilibrium is disrupted by quantum-tunnelling nucleation of
bubbles of the new vacuum state, and these bubbles eventually expand
and carry the phase transition to all space. 
CP is violated during the periods of phase change as the Higgs field
is changing.
Within the minimal Standard Model
 it is not possible to achieve the fundamental
criterion
$\frac{v}{T_{c}} > 1$.
The MSSM can satisfy the condition,
though there are tight theoretical constraints:
$M_h < 120$~GeV and $m_{stop} < m_{top}$ for $\tan{\beta} > 5$~\cite{barmssm},
however others are less optimistic~\cite{prokopec}.
The precision measurements from a linear collider
 of the stop mass and mixing parameters
are essential for probing electroweak baryogenesis.
The CP contraints would be obtained from precision measurement of
chargino and neutralino masses and mixings.
 
There are other mechanisms for baryogenesis worth mentioning.
In leptogenesis~\cite{leptogen}
%
%
%
%
%
%
a lepton asymmetry creates a baryon asymmetry 
via EW processes. The Sakharov criteria are satisfied by the decay of
a heavy lepton which has frozen out of thermal equilibrium.
Such heavy leptons could be, e.g., a right-handed Majorana neutrino
which violates the B-minus-L conservation.
The SM can be extended to accomodate such particle by embedding in 
the GUT group SO(10), which is already an attractive way to describe
the smallness of neutrino masses.
The linear collider could shed light on this scenario
by accessing GUT-scale physics via precision electroweak measurements
or by direct observation of lepton flavor violation.
Affleck and Dine~\cite{affleckdine} have suggested a scheme for
%
%
baryogenesis from the decay of scalar cosmological fields carrying
nonzero baryon number. The MSSM facilitates the mechanism nicely,
providing yet another theoretical scenario requiring precision
measurements in the electroweak sector.

\subsection{Dark Energy}

The fact that most of the energy density in the universe is neither in
the form of matter or radiation is undeniable from the case made with
data from the Type Ia supernova survey, fits to the CMB anisotropy,
and the clustering properties of radio galaxies.
The source of this dark energy remains a mystery.
Some possible explanations are
exotic supersymmetry breaking,
extra-dimensions,
scalar fields with both kinetic and potential terms in the Lagrangian,
and modifications to gravity.
Of these, the first three might be resolved with linear collider data.

If supersymmetry broken at the 
electroweak scale  were the correct theory, the dark energy
 should be at the supersymmetry
breaking energy scale of 1~TeV. The vacuum energy density is
$\rho_{vac} = \mathrm{E}_{vac}^{4} = (10^{-3}~eV)^{4}$,
so the theory differs with observation by 15 orders of
 magnitude~\cite{astrorev}.
But perhaps supersymmetry is broken differently. 
Discovering supersymmetry
 and performing precision measurements leading to a
understanding of the breaking might lead to understanding of the dark
energy.

In some models with extra dimensions,
the spatial geometry contributes to the apparent 3D vacuum energy
significantly. 
For instance, in brane models with two extra dimensions,
the brane tension affects the observed cosmological constant.
If the extra dimensions are sufficiently large,
a linear collider could shed light on the dark energy problem
by mapping out the spectrum of Kaluza-Klein excitations.

The dynamical scalar field in models like quintessence~\cite{0302151}
interacts with matter.
The dark energy can therefore interact with the dark matter,
and if the dark matter is in the form of, e.g., neutralinos,
measurements of the supersymmetric
particle masses might again shed light on the dark
energy.

\clearpage

\section{Conclusions}
\label{conclusions}

We hope that this review has served as an (all too brief!) introduction
to {\it some} of the exciting physics potentially accessible at a 
500~GeV -- 1~TeV linear electron positron collider.  
The technology now exists to build a
machine of the required energy and luminosity to explore 
{\it with high precision}
the energy range pertinent to the electroweak theory.
It has been argued that the high precision measurements
play a role both orthogonal to the measurements the LHC will make
{\it and} supportive of some of the physics LHC will possibly see.
For this reason  a linear collider should be operational before
the end of LHC running.  
Precision measurements at the TeV electroweak scale have been shown to
be a potential window on much higher energy scales (even the Planck
scale) in a large class of theories.

Within the electroweak theory, an electron-positron collider has been shown to have
the capability to pin down quantum numbers and explore the chiral
nature of couplings by virtue of the superior energy resolution and
polarized beams. Options for collisions of electrons or photons have
been described which shed even more light on the nature of couplings
and can enhance signals over backgrounds.
Specific examples have been given of the power 
a linear collider has to study the  Higgs sector and 
identify particles and measure parameters in supersymmetric models. 

Only a sampling of the alternative models have been described.
Some of these theories could account for the large energy gap between
the Planck and electroweak scales.
Others provide possible explanations for the cosmological dark matter
or dark energy.  
While we cannot predict exactly what new physics will emerge from the
measurements at the TeV scale, we know from virtual effects in recent
measurements that this energy scale will be active and lead to better
understanding of the electroweak theory ... and possibly beyond.
We argue that the high energy reach, particle production,
precision and polarization  of a combined LHC
and  $\mathrm{e}^{+}\mathrm{e}^{-}$ collider program will be necessary to
fully explore this physics.

\section{Acknowledgements}
MO is supported in part by the National Science Foundation
under award NSF-PHY/0201792.
SD is supported by the Department of Energy, Contract
DE-AC02-76CH00016.
The authors thank the following for their generous assistance in
preparing this article: 
 Uli Baur, Jim Brau, Sean Carroll,  
Joe Lykken,  Michael Peskin,
Chris Quigg,
David Rainwater,  
Tor Raubenheimer, Geraldine Servant, Savdeep Sethi, Mark Trodden, Michael Woods.
\section{Bibliography}


\begin{thebibliography}{99}


%
\bibitem{SM} 
Weinberg S. {\it  Phys. Rev. Lett.} 19:1264 (1967);
Salam A. In {\it Elementary~ Particle~ Theory}.
Stockholm: Almquist and Wiksells (1969);
Glashow SL, Iliopoulos J, Maiani L. {\it Phys.~ Rev.~ D} 2:1285
(1970)

\bibitem{fine}
't Hooft G. In {\it Recent~ Developments~ in~
 Gauge~ Theories},  ed
 G. 't Hooft, et al., p. 135. New York:Plenum (1980);
Witten E. {\it Nucl. Phys. B} 188:513
(1981); Susskind L. {\it Phys. Rep.} 104:1
(1984)

\bibitem{pt}
Peskin M, Takeuchi T. {\it Phys. Rev. Lett.}
65:964 (1990); {\it Phys. Rev. D}
46:381 (1992)

\bibitem{LEPEW}
LEP and SLD Collaborations. hep-ex/0312023

\bibitem{LEPEWWG}
LEP Electroweak Working Group website.
http://www.cern.ch/LEPEWWG 

\bibitem{LEPHiggs}
LEP Collaborations, Phys. Lett. {\bf B565}, 61 (2003),
hep-ex/0306033

\bibitem{lqt}
Lee B, Quigg C, Thacker H.
 {\it Phys. Rev. D} 16:1519 (1977)

\bibitem{lhc}
Branson J, et al. (ATLAS and CMS
Collabs.). {\it Eur. Phys. J. direct C}
4:N1 (2002);
Baur U, et al. {\it Proc.~ Snowmass~ Study~ on~ the~ Future~
of~ Particle~ Physics}, Snowmass, Colorado,
30 June-21 July 2001, (2001), hep-ph/0201227

\bibitem{atlastdr}
ATLAS Collaboration.
ATLAS Physics Technical Design Report,
http://atlas.web.cern.ch/Atlas/GROUPS/PHYSICS/TDR/TDR.html;
CMS Collaboration. 
CERN-LHCC-94-38 (1994)


\bibitem{HEPAP}
High Energy Physics Advisory Panel. 
DOE/SC-0027 (2000);
U.S. Department of Energy. ``High Energy Physics Facilities 
Recommended for the Department of Energy Office of Science
Twenty-year Roadmap'' (2003) 
(http>//doe-hep.hep.net/HEPFacSub/HEPAP\_FacilitiesMar03.pdf)

\bibitem{pm}
Peskin M, Murayama H. {\it Ann.~ Rev.~ Nucl.~ Part.~ Sci.} 
46:533 (1996), hep-ex/9606003

\bibitem{accomando}
Accomando E, et al. {\it Phys. Rept.}
299:1 (1998),
hep-ph/9705442

\bibitem{ellis0210052}
Ellis J. hep-ex/0210052

\bibitem{concensus}
Grannis P, ed.,
{\it Understanding~ Matter,~ Energy,~ Space~ and ~Time:~ The~
 Case~ for~ the~Linear~ Collider},
http://sbhep1.physics.sunysb.edu/$\sim$grannis/ilcsc/lc\_concensus.pdf;
ALCWG, FERMILAB-PUB-00/152 (2000)

\bibitem{snow2001} 
The American Linear Collider Working Group,
{\em Linear Collider Physics Resource Book for 
Snowmass 2001}, May 2001, BNL-52627, CLNS 01/1729,
FERMILAB-Pub-01/058-E, LBNL-47813, SLAC-R-570,
UCRL-ID-143810-DR
(http://www.slac.stanford.edu/grp/th/LCBook).
Updates and additions to this document in the
Snowmass 2001 eProceedings,\\ 
http://www.slac.stanford.edu/econf/C010630/proceedings.shtml

\bibitem{tesla}
{\em TESLA Technical Design Report}, 
March 2001, DESY 2001-11 
(http://www.desy.de/$\sim$lcnotes/tdr)

\bibitem{acfa} 
ACFA Linear Collider Working Group.
{\em ACFA Linear Collider Working Group Report},
November 2001, KEK-Report 2001-11, hep-ph/0109166

\bibitem{bagger} 
Bagger J, et al. 
hep-ex/0007022




\bibitem{simmons}
Simmons E. 
hep-ph/0011244

\bibitem{hills}
Hill C, Simmons E. {\it Phys. Rept.}
381:235 (2003), 
hep-ph/0203079

\bibitem{littleh}
Arkani-Hamed N, Cohen AG, Georgi H.
{\it Phys. Lett. B} 513:232 (2001),
hep-ph/0105239;
Arkani-Hamed N, Cohen AG, Katz E,
Nelson A. {\it JHEP} 07:034 (2002),
hep-ph/0206021;
Arkani-Hamed N, Cohen AG, Gregoire T, 
Wacker JG. {\it JHEP} 08:020 (2002),
hep-ph/0202089

\bibitem{nohiggs}
Csaki C, Grojean C, Murayama H, Pilo L, 
Terning J. hep-ph/0305237;
Hall LJ, Nomura Y. {\it Phys. Rev. D}
64:055003 (2001),
hep-ph/0103125;
Davoudiasl H, Hewett J, Rizzo T. 
hep-ph/0312193 



\bibitem{astrorev}
Trodden M, Carroll S. In
{\it Proc.~Theoretical~ Advanced~ Study~Institute~
in~ Elementary~Theory}, Boulder, Colorado, 1-27 June
2003, astro-ph/0401547

\bibitem{susydm}
Baer H, Balazs C, Belyaev A, O'Farrill J.
{\it JCAP} 0309:007 (2003),
hep-ph/0305191;
Ellis J, Feng J, Ferstl A, Olive K.
{\it Eur. Phys. J. C} 24:311 (2002),
astro-ph/0110225; 
Feng J, Matchev K, Wilczek F.
{\it Phys. Rev. D} 63:045024 (2001),
astro-ph/0008115

\bibitem{eoss}
Ellis J, Olive K, Santoso Y, Spanos VC. 
{\it Phys. Lett. B} 565:176 (2003), 
hep-ph/0303043

\bibitem{jo}
Hewett J, Spiropulu M.
{\it Ann.~Rev.~Nucl.~Part.~Sci} 52:396 (2002),
hep-ph/0205106

\bibitem{gw}
Weiglein G.
 LHC/LC Study group report,
http://www.ippp.dur.ac.uk/~georg/lhclc/

\bibitem{comp}
Desch K, Kalinowski J, Moortgat-Pick G, Nojiri M,
 Polesello G.
 hep-ph/0312069

\bibitem{scopeA}
Oreglia M. ed. {\it Design~ Considerations~
 for~ an~ International~ Linear~
Collider}, http://blueox.uoregon.edu/$\sim$lc/scope.ps (2003)

\bibitem{scopeI}
Heuer R. ed. {\it Parameters~ for~ the~ Linear~ Collider},
http://www.fnal.gov/directorate/icfa/LC\_parameters.pdf (2003)

\bibitem{batt1}
Battaglia M.
In {\it Proc.~  10$^{th}$~
International~ Conference~ on ~Supersymmetry~
and~ Unification~ of~ Fundamental~ Interactions},
Hamburg, Germany, June 17-23 2002, hep-ph/0211461

\bibitem{kd} 
Desch K.
{\it Proc.~ 4$^{th}$~ ECFA/DESY~ Workshop~ on~ Physics~
and~ Detectors~ for~ a ~90~ GeV~ to~ 800~
 GeV~ Linear~ $e^+e^-$~ Collider},
Amsterdam, The Netherlands, 1-4 Apr. 2003,
hep-ph/0311092 


\bibitem{bd}
Battaglia M, Desch K. 
{\it Proc. ~5$^{th}$~ International~
Linear~Collider~Workshop},
Batavia, Ill., 24-28 Oct. 2000, hep-ph/0101165

\bibitem{ghk}
Gunion J, Haber H, Van~Kooten R.
In
{\it Linear~ Collider~Physics~in ~the~New~
Millennium}, ed. K.~Fujii, et al.,
Singapore: World Scientific,  hep-ph/0301023
 

\bibitem{zerwas}
Behnke T, Wells J, Zerwas P.
 {\it Prog. Part. Nucl. Phys.} 48:363 (2002)

\bibitem{bdr}
Battaglia M, De Roeck A.  
{\it Proc.~ Snowmass~ Study~ on~ the~ Future~
of~ Particle~ Physics}, Snowmass, Colorado,
30 June-21 July (2001), hep-ph/0111307


\bibitem{erler}
Erler J, et al.
{\it Proc.~ Snowmass~ Study~ on~ the~ Future~
of~ Particle~ Physics}, Snowmass, Colorado,
30 June-21 July (2001), hep-ph/0112070

\bibitem{accel}
Edwards DA, Syphers MJ.
{\it An~ introduction~ to ~the~
 Physics ~of~ High~ Energy~ Accelerators} Chicago:Willey (1993);
Chao AW, Tigner M.
{\it Handbook~ of~ Accelerator~ Physics~
 and ~Engineering}, Singapore:World Scientific (1999)

\bibitem{heuer}
Heuer R. {\it Int. Jour. Mod. Phys. A} 17:3469 (2002),
hep-ex/0111070

\bibitem{warmRF}
Phinney N, et al. (NLC Collab.).
SLAC-R-571 (2001)

\bibitem{jlcref}
 JLC Collab, KEK report 97-1, http://acfahep.kek.jp/acfareport/index.html

\bibitem{CLIC}
 Guignard  G, et al. (CLIC Study Team).
 CERN 2000-008 (2000)

\bibitem{ILCTRC}
Loew G, et al.
(International LC Technical Review Committee)
http://www.slac.stanford.edu/xorg/ilc-trc/ilc-trchome.html

\bibitem{ILCTRP}
International Linear Collider Steering Committee website:  
http://www.fnal.gov/directorate/icfa/International\_ILCSC.html 

\bibitem{beamstrahlung}
Chen P. Phys. Rev. {\bf D} 46:1186 (1992);
%
Schulte D.
eConf C980914:127 (1998)

\bibitem{IPBI}
Cinabro D, Torrence E, Woods M.
{\it Status~ of~ LC~ Beam ~Instrumentation~ Design},
ALCPG Note  LCD-ALCPG-03-0001 (2003),
http://www.slac.stanford.edu/xorg/lcd/ipbi/notes/white.pdf;
%
Frary MN, Miller DJ.
In *Munich/Annecy/Hamburg 1991, Proceedings, e+ e- collisions at 500-GeV,
part A, 379-391 (1981);
%
Boogert ST, Miller DJ.
To appear in the proceedings of International Workshop on Linear Colliders
(LCWS 2002), Jeju Island, Korea, 26-30 Aug 2002, hep-ex/0211021;
%


\bibitem{polarimeter}
 Woods M.
{\it Int.~ J. ~Mod.~ Phys.~  A} 15:2529 (2000), 
hep-ex/0004004

\bibitem{barger}
Barger V, Beacom JF, Cheung K, and Han T,
Phys. Rev. {\bf D50}:6704 (1994),
hep-ph/9404335

\bibitem{fp}
Feng J, Peskin M.
{\it Phys. Rev. D } 64:115002
(2001),
hep-ph/0105100

\bibitem{bfmpp}
Blochinger C, Fraas H, Moortgat-Pick G, Porod W.
hep-ph/0201282

\bibitem{fmz}
Freitas A, von~Manteuffel A, Zerwas P.
 hep-ph/0310182

\bibitem{gamgam}
Gunion J.
 {\it Int. J.~Mod. Phys. A} 18:2739 (2003); 
Asner D, et al. hep-ph/0308103 

\bibitem{zfac}
Rowson PC, Woods M.
hep-ex/0012055

\bibitem{giga}
Erler J, Heinemeyer S, Hollik W,
Weiglein G, Zerwas P.
{\it Phys. Lett. B} 486:125 (2000),
hep-ph/0005024; 
Baur U, et al. {\it Proc.~ Snowmass~ Study~ on~ the~ Future~
of~ Particle~ Physics}, Snowmass, Colorado,
30 June-21 July 2001, hep-ph/0202001

\bibitem{schummdet}
Schumm BA, talk presented at the 2001 Snowmass Workshop on the Future
of Particle Physics (2001),
hep-ex/0111009.

\bibitem{batdet}
Battaglia M, Hinchliffe I, Jaros J, and Wells J,
ibid.,
hep-ex/0201018

\bibitem{tesla_det}
TESLA Collaboration, in Part IV of
the TESLA Technical Design Report,
DESY/2001-011 (2001)
 

\bibitem{brau_sid}
Brau JE. In 
{\it Proc.
Extended Joint ECFA/DESY Study on Physics and Detectors for a Linear 
Electron-Positron Collider}, ed. R. Settles (2003),
http://www.desy.de/conferences/ecfa-desy-lcext.html

\bibitem{videau}
Brient JC, Videau H,
hep-ex/0202004

\bibitem{brau_det}
 Brau JE, In 
{\it Proc. of the International Workshop on Linear Colliders 2002},
p. 675, ed.  J.S. Kang and S.K. Oh, Korean Physical Society (2003) 




 



\bibitem{zep1}
Zeppenfeld D.
 {\it Proc.~ Snowmass~ Study~ on~ the~ Future~
of~ Particle~ Physics}, Snowmass, Colorado,
30 June-21 July 2001, hep-ph/0203123;
M.~Duerhson.
 ATLAS Physics Note (2003)


\bibitem{ctw}
Choudhury D, Tait T, Wagner, CEM.
{\it Phys. Rev. D} 65:115007 (2002),
hep-ph/0202162

\bibitem{kniehl}
Kniehl BA. hep-ph/0210175;
Dittmaier S. hep-ph/0308079

\bibitem{brau_higgs}
Potter C, Brau J, Iwasaki M.  
 {\it Proc.~ Snowmass~ Study~ on~ the~ Future~
of~ Particle~ Physics}, Snowmass, Colorado,
30 June-21 July (2001)


\bibitem{tesla_higgs}
Kuhl J.  http://www-flc.desy.de/ecfa-higgs/montpellier/kuhl-montpellier.pdf

\bibitem{juste}
Juste A, Merino G. hep-ph/9910301;
Gay A. ECFA/DESY Workshop;
Baer H, Dawson S, Reina L.
{\it Phys. Rev. D} 61:013002 (2000),
hep-ph/9906419

\bibitem{higgpot}
Baur U, Plehn T, Rainwater D.
 {\it Phys. Rev.  D} 68:033001 (2003),
hep-ph/0304015;
Castanier C, Gay P, Lutz P, Orloff J.  hep-ph/0101028;
Battaglia M, Boos E,  Yao WM. hep-ph/0111276

\bibitem{lam4}
Djouadi A, Kilian W, M\"uhlleitner MM,  Zerwas P.
{\it Eur. Phys. J. C} 10:27 (1999),
hep-ph/9903229

\bibitem{miller}
 Miller DJ, Choi SY, Eberle B,
 M\"uhlleitner MM, Zerwas P.
 {\it Phys. Lett. B} 505:149 (2001),
hep-ph/0102023

\bibitem{dgl}
Dova MT, Garcia-Abia P, Lohmann W.
hep-ph/0302113

\bibitem{cmmz}
Choi SY, Miller DJ, M\"uhlleitner MM, Zerwas P.
{\it Phys. Lett. B} 553:61 (2003),
hep-ph/0210077;
{\it Phys. Rev. D} 49:79 (1994),
hep-ph/9306270


\bibitem{mhmax}
Degrassi G, Heinemeyer S, Hollik W, Slavich P, 
Weiglein G. {\it Eur. Phys. J. C} 28:133 (2003),
hep-ph/0212020

\bibitem{nmssm}
Quiros M, Espinosa J. In
{\it Particles, Strings, and Cosmology}, Boston (1998), hep-ph/9809269;
Kane G, Kolda C, Wells J.  {\it Phys. Rev.
Lett.} 70:2686 (1993), hep-ph/9210242.


\bibitem{guasch}
Guasch J, Hollik W,  Penaranda S. hep-ph/0307012;
{\it Phys. Lett. B} 515:367 
(2001), hep-ph/0106027

\bibitem{batt}
Battaglia M, Desch K. hep-ph/0102265 

\bibitem{single}
Moretti S. hep-ph/0306297;
hep-ph/0209210;
Logan H,  Su SF.
 {\it Phys. Rev. D} 66:035001 (2002), 
hep-ph/0203270



\bibitem{suprev}
Martin  SP.
 {\it A~Supersymmetric~Primer}, In {\it
Perspectives ~on~Supersymmetry}, ed. G. Kane
Singapore: World
Scientific (1997),
hep-ph/9709356; 
Chung D, Everett L, Kane G, King S, Lykken J.
hep-ph/0312378;
Dawson S. 
{\it Proc.~Theoretical~ Advanced~ Study~ Institute~ in~
Elementary~ Particles}, Boulder, Colorodo, June 1997,
hep-ph/9712464;
Polonsky N.  {\it Lecture Notes in Physics Monographs},
Vol. m68.  Heidelberg:  Springer-Verlag (2001),
hep-ph/0108236

\bibitem{fn}
Feng J, Nojiri M. hep-ph/0210390 

\bibitem{snowsusy}
Danielson MN, et al. 
{\it Proc.~ Snowmass~ Study~ on~ the~ Future~
of~ Particle~ Physics}, Snowmass, Colorado,
30 June-21 July 2001

\bibitem{drees}
Drees M.
{\it  Lectures ~given~ at~ the~ Inauguration~ Conference~
of~ the~ Asia~ Pacific~ Center~ for ~Theoretical~ Physics},
Seoul, Korea, 4-19 June 1996, hep-ph/9611409

\bibitem{bhp}
Bachacou H, Hinchliffe I, Paige F.
{\it Phys. Rev. D} 62:015009 (2000),
hep-ph/9907518

\bibitem{snow}
Battaglia M, et al.,
{\it Proc.~ Snowmass~ Study~ on~ the~ Future~
of~ Particle~ Physics}, Snowmass, Colorado,
30 June-21 July 2001,
hep-ph/0201177

\bibitem{dre}
Dreiner H.
 {\it An~Introduction~to ~Explicit ~R~Parity
~Violation,
Perspectives ~on~Supersymmetry}, ed. G. Kane
Singapore: World
Scientific (1997),
hep-ph/9707435

\bibitem{fmpt}
Feng J, Peskin M, Murayama H, Tata X.
 {\it Phys. Rev. D}
52:1418 (1995),
hep-ph/9502260

\bibitem{mb}
Martyn H, Blair G. hep-ph/9910416

\bibitem{msugra}
Chamseddine A, Arnowitt R, Nath P.
 {\it Phys. Rev. Lett.} 49:970 (1982);
Barbieri R, Ferrara S, Savoy C.
{\it Phys. Lett. B} 119:343 (1982);
Hall L, Lykken J, Weinberg S.
{\it Phys. Rev. D} 27:2359 (1983) 

\bibitem{dd}
Djouadi A, Drees M, Kneur J.
{\it JHEP} 0108:055 (2001), 
hep-ph/0107316

\bibitem{lcsusy}
Baer H, Belyaev A, Krupovnickas T, Tata X.
hep-ph/0311351

\bibitem{gauge_med}
Dine M, Fischler W. {\it Nucl. Phys. B} 204:346 (1982);
Dimopoulos S, Raby S. {\it Nucl. Phys. B} 219:479 (1983);
Dine M, Nelson A. {\it Phys. Rev. D} 48:1277 (1993), hep-ph/9303230;
Dine M, Nelson A, Shirman Y.  {\it Phys. Rev. D} 51:1362 (1995)

\bibitem{unig}
Blair G, Porod W, Zerwas P.
 {\it Phys. Rev. D} 63:017703 (2001),
hep-ph/0007107;
{\it Eur. Phys. J. C} 27:263 (2003),
hep-ph/0210058

\bibitem{bfs}
Bagger J, Falk A, Schwartz M.
 {\it Phys. Rev. Lett.} 84:1385 (2000),
hep-ph/9908327;
Chivukula RS, Hoelbling C,  Evans N.
 {\it Phys. Rev. Lett.} 85:511 (2000),
hep-ph/0002022;
Alam S, Dawson S, Szalapski R.
 {\it Phys. Rev. D} 57:1557 (1998),
hep-ph/9706542;
Barbieri R, Strumia A
 {\it Phys. Lett. B} 462:144 (1999),
hep-ph/9905281;
Kolda C, Murayama H.
 {\it JHEP} 00007:035 ( 2000),
hep-ph/0003170


\bibitem{barklow}
Barklow T, Chivukula RS, Goldstein J, Han T.
{\it Proc.~ Snowmass~ Study~ on~ the~ Future~
of~ Particle~ Physics}, Snowmass, Colorado,
30 June-21 July 2001, hep-ph/0201243

\bibitem{pw}
Peskin M, Wells J.
{\it Phys. Rev.  D} 64:093003 (2001), 
hep-ph/0101342

\bibitem{fms}
Fujii K, Matsui T, Sumino Y.
 {\it Phys. Rev. D} 50:4341 (1994)


\bibitem{mm}
Martinez M, Miquel R.
{\it Eur. Phys. Jour. C} 27:49 (2003), 
hep-ph/0207315

\bibitem{hoang}
Hoang AH, hep-ph/0307376

\bibitem{tmass}
Hoang AH, Teubner T.
 {\it Phys. Rev. D} 60:114027 (1999),
hep-ph/9904468;
 Hoang AH, Ligeti Z, Manohar  AV. 
{\it Phys. Rev. Lett.} 82:277 (1999),
hep-ph/9809423

\bibitem{hoang2} 
Hoang AH, Stewart IW.
 {\it Phys. Rev. D} 67:114020 (2003),
hep-ph/0209340




\bibitem{ADD}
Arkani-Hamed N, Dimopoulos S, Dvali G.
{\it Phys. Lett.  B} 429:263 (1998), 
hep-ph/9803315

\bibitem{0007226}
Antoniadis I, Benakli K.
{\it Int. J. Mod. Phys. A} 15:4237 (2000),
hep-ph/0007226

\bibitem{optns}
Davoudiasl H.
{\it Phys.Rev. D} 61:044018 (2000), 
hep-ph/9907347;
Ghosh DK, Raychaudhuri S.
{\it Phys.Lett. B} 495:114 (2000),
hep-ph/0007354;
Choudhury SR, Cornell AS, Joshi GC.
{\it Phys.Rev. D} 64:114022 (2001), 
hep-ph/0105002;
{\it Phys.Lett. B} 535:289 (2002), 
hep-ph/0202272;
Ghosh DK, Mathews P, Poulose P, Sridhar K.
{\it JHEP} 9911:004 (1999), 
hep-ph/9909567

\bibitem{0211374}
Rizzo T. 
{\it JHEP} 0302:008 (2003),
hep-ph/0211374

\bibitem{0103053}
\`Eboli OJ, Magro MB, Mathews P, Mercadante PG.
{\it Phys. Rev.  D} 64:035005 (2001),
hep-ph/0103053

\bibitem{9811291}
Giudice GF, Rattazzi R, Wells J.
{\it Nucl.Phys.  B} 544:3 (1999), 
hep-ph/9811291

\bibitem{9811350}
Han T, Lykken J, Zhang RJ.
{\it Phys. Rev. D} 59:105006 (1999), 
hep-ph/9811350

\bibitem{0110346}
Cheung K, Landsberg G.
{\it Phys. Rev. D} 65:076003 (2002), 
hep-ph/0110346

\bibitem{9902263}
Agashe K, Deshpande NG.
{\it Phys. Lett.  B} 456:60 (1999), 
hep-ph/9902263

\bibitem{0010354}
Hewett J, Petriello FJ,  Rizzo T.
{\it Phys. Rev.  D} 64:075012 (2001),
hep-ph/0010354

\bibitem{9811337}
Mirabelli EA, Perelstein M,  Peskin M.
{\it Phys. Rev. Lett.} 82:2236 (1999),
hep-ph/9811337

\bibitem{0110339}
Gopalakrishna S, Perelstein M,  Wells J.
hep-ph/0110339

\bibitem{0307117}
Dutta S, Konar P, Mukhopadhyaya B,  Raychaudhuri S.
{\it Phys. Rev.  D} 68:095005 (2003),
hep-ph/0307117


\bibitem{rs}
Randall L, Sundrum R.
 {\it Phys. Rev. Lett.} 83:4690
(1999),
hep-th/9906064; 
{\it Phys. Rev. Lett.}
83:3370 (1999),
hep-ph/9905221

\bibitem{0006041}
Davoudiasl H, Hewett J, Rizzo T.
{\it Phys. Rev. D} 63:075004 (2001),
hep-ph/0006041

\bibitem{0307096}
Rai SK, Raychaudhuri S.
{\it JHEP} 0310:020 (2003),
hep-ph/0307096

\bibitem{rad}
Hewett J, Rizzo T.
{\it JHEP} 0308:028
(2003), 
[hep-ph/0202155].


\bibitem{0012100}
Appelquist T, Cheng HC, Dobrescu B.
{\it Phys. Rev.  D} 64:035002 (2001),
hep-ph/0012100

\bibitem{0312055}
Carone CD, Conroy JM, Sher M, Turan I.
hep-ph/0312055



%

\bibitem{lindner}
Lindner EV, ed. 
{\it Resource~ Book~ on~ Dark~ Energy},
compilation from Snowmass 2001,
http://supernova.lbl.gov/~evlinder/sci.html

\bibitem{HuURL}
http://background.uchicago.edu/$\sim$whu/physics/physics.html 

\bibitem{perlmutter3} 
 Perlmutter S, et al. (Supernova Cosmology 
Project Collab.). 
{\it Astrophys. Journ.} 517:565
(1999), 
astro-ph/9812133

\bibitem{bennett}
Bennett CL,   et al. 
{\it Astrophys. J. Suppl. } 148:1 (2003),
astro-ph/0302207

\bibitem{spergel}
Spergel DN, et al. 
{\it Astrophys. J. Suppl.} 148:175 (2003), 
astro-ph/0302209

\bibitem{ccn}
Chattopadhyay U, Corsetti A, Nath P. 
{\it Phys. Rev. D} 68:035005 (2003), 
hep-ph/0303201

\bibitem{verde}
Verde L.
{\it  New ~Astronomy~ Reviews}  47:713 (2003)

\bibitem{jgk}
Jungman G, Kamionkowski M, Griest K.
{\it  Phys. Rept. } 267:195 (1996),
hep-ph/9506380


\bibitem{feng0012277}
Feng J. hep-ph/0012277


\bibitem{strongCP}
Cheung HY.
{\it Phys. Rept. } 158:1 (1988);
Kim JE.
{\it  Phys. Rept.} 150:1 (1987)

\bibitem{0206071}
Servant G, Tait T.
{\it  Nucl. Phys.  B} 650:391 (2003),
hep-ph/0206071

\bibitem{0302041}
Cembranos JA, Dobado A, Maroto AL. 
{\it Phys. Rev. Lett.} 90:241301 (2003),
hep-ph/0302041

\bibitem{CFM}
Cheng HC, Feng J, Matchev KT.
{\it Phys. Rev. Lett.} 89:211301 (2002), 
hep-ph/0207125

\bibitem{EFGO}
Ellis J, Falk T, Ganis G, Olive K.
{\it Phys. Rev. D} 62:075010 (2000),
hep-ph/0004169

\bibitem{birkedal}
Birkedahl-Hansen A.
 private communication


\bibitem{fr}
 Richard F.
{\it Proc. 21$^{st}$ International
Symposium on Lepton and Photon Interactions at
High Energies}, Batavia, Ill., 11-16 Aug. 2003,
hep-ph/0312020

\bibitem{sakarov}
Sakharov AD.
{\it Zh. Eksp. Teor. Fiz. Pis'ma} 5:32 (1967);
{\it  JETP Lett. B} 91:24 (1967)

\bibitem{GUTBAU}
Langacker P.
{\it Phys. Rept.} 72:185 (1981)

\bibitem{0302151}
Trodden M. hep-ph/0302151,
Zlatev I, Wang L, Steinhardt P.
{\it Phys. Rev. Lett.} 82:896 (1999), astro-ph/9807002

\bibitem{barmssm}
Carena M, Quiros M, Seco M, Wagner CEM, {\it Nucl. Phys. B} 650:24 (2003),
hep-ph/0208043

\bibitem{prokopec}
Procopec T et al. hep-ph/0302192

\bibitem{leptogen} 
Mohapatra RN, Senjanovic G.
{\it  Phys. Rev.  D} 12:1502 (1975);
 Fukugita  M, Yanagida T.
{\it  Phys. Lett. } 174:45 (1986);
Gell-Mann M, Ramond P, Slansky R.
In {\it Supergravity}, 
ed. P. Van Nieuwenhuizen and D.Z. Freedman, North Holland (1979);
Langacker P, Peccei RD, Yanagida T,
{\it  Mod. Phys. Lett. A} 1:541 (1986);
Luty M.
{\it Phys. Rev.  D} 45:455 (1992)

\bibitem{affleckdine}
Affleck I, Dine M.
{\it  Nucl. Phys.  B} 249:361 (1985)







\end{thebibliography}
\end{document}